\documentclass{elsart3}
\usepackage{graphics}
\usepackage{epsfig}
\usepackage{amssymb}
\usepackage{cite}
\usepackage{subfigure}

\begin{document}

\newcommand{\greeksym}[1]{{\usefont{U}{psy}{m}{n}#1}}
\newcommand{\umu}{\mbox{\greeksym{m}}}
\newcommand{\udelta}{\mbox{\greeksym{d}}}
\newcommand{\uDelta}{\mbox{\greeksym{D}}}
\newcommand{\uOmega}{\mbox{\greeksym{W}}}
\newcommand{\uPi}{\mbox{\greeksym{P}}}
\newcommand{\ualpha}{\mbox{\greeksym{a}}}
 
\newcommand{\mrm}{\mathrm}
\newcommand{\Neq}{\mrm{n}_{\mrm{eq}}/\mrm{cm}^{2}}

\begin{frontmatter}




\title{Design and performance of the silicon sensors \\ for the CMS barrel pixel detector}

\author[uniz]{Y.~Allkofer},
\author[uniz]{C.~Amsler},
\author[purdue]{D.~Bortoletto},
\author[uniz]{V.~Chiochia\corauthref{cor1}}, \ead{vincenzo.chiochia@cern.ch}
\author[miss]{L.~Cremaldi},
\author[basel]{S.~Cucciarelli},
\author[uniz,psi]{A.~Dorokhov},
\author[uniz,psi]{C.~H\"ormann},
\author[psi]{R.Horisberger},
\author[jhu]{D.~Kim},
\author[warsaw]{M.~Konecki},
\author[psi]{D.~Kotlinski},
\author[uniz,psi]{K.~Prokofiev},
\author[uniz]{C.~Regenfus},
\author[psi]{T.~Rohe\corauthref{cor1}},\ead{tilman.rohe@psi.ch}
\author[miss]{D.~A.~Sanders},
\author[purdue]{S.~Son},
\author[jhu]{M.~Swartz},
\author[uniz]{T.~Speer}

\corauth[cor1]{Corresponding authors}
\address[uniz]{Physik Institut der Universit\"at Z\"urich-Irchel, 8057 Z\"urich, Switzerland}
\address[purdue]{Purdue University, West Lafayette, IN 47907, USA}
\address[miss]{University of Mississippi, University, MS 38677, USA}
\address[basel]{Institut f\"ur Physik der Universit\"at Basel, 4056 Basel, Switzerland}
\address[jhu]{Johns Hopkins University, Baltimore, MD 21218, USA}
\address[warsaw]{Institute of Experimental Physics, University of Warsaw, Warsaw, Poland}
\address[psi]{Paul Scherrer Institut, 5232 Villigen PSI, Switzerland}
\begin{abstract}
The CMS experiment at the LHC includes a hybrid silicon pixel detector
for the reconstruction of charged tracks and of the interaction vertices.
The barrel region consists of n-in-n sensors with $100\times150~\mu$m$^2$ cell
size processed on diffusion oxygenated float zone silicon. A biasing grid is
implemented and pixel isolation is achieved with the moderated p-spray technique.
An extensive test program was carried out on the H2 beam line of the CERN SPS.
In this paper we describe the sensor layout, the beam test setup  
and the results obtained with both irradiated and non-irradiated prototype devices. 
Measurements of charge collection, hit detection efficiency, Lorentz angle 
and spatial resolution are presented. 
\end{abstract}
\end{frontmatter}

\section{Introduction\label{sec:introduction}}


The CMS experiment, currently under construction at the Large Hadron Collider
(LHC) will include a silicon pixel detector~\cite{CMSTrackerTDR:1998} to allow tracking in the region closest
to the interaction point. The detector will be a key component for re\-con\-struc\-ting interaction
vertices and heavy quark decays in a particularly harsh environment, characterized by
a high track multiplicity and heavy irradiation.
The detector will consist of three barrel layers
and two disks at each end of the barrel. The innermost barrel layer 
has a radius of 4.3 cm, while for the second and third layers the radius is
of 7.2 cm and 11 cm, respectively. The layers are composed of modular detector units.
These modules consist of thin, segmented silicon sensors with highly integrated readout
chips connected by the bump bonding technique.
They are attached to cooling frames, the cooling tubes being an integral part
of the mechanical structure. 
The minimal pixel size is dictated by the readout circuit area required for
each pixel. In finding and localizing secondary decay vertices both transverse ($r\phi$)
and longitudinal ($z$) coordinates are important in the barrel region. 
Therefore a nearly square pixel shape is preferred. Because charge is often shared 
among several pixels, the use of analogue signal readout enables position interpolation improving the spatial resolution.
In the barrel the charge sharing in the $r\phi$-direction will
be large due to the 4 T magnetic field of CMS. With a sensitive detector thickness of 285 $\mu$m 
the pixel size will be 100 $\mu$m and 150 $\mu$m along the $r\phi$ and $z$
coordinates, respectively. These value will give favourable spatial resolution and
cluster sizes. 

One of the greatest challenges in the design of the pixel detector is the high 
radiation level on all components at very close distances to the colliding beams.
At full LHC luminosity the innermost layer will be exposed to a 
particle fluence of $3\times10^{14}$ n$_{\rm{eq}}/$cm$^2$/yr~\footnote{All fluences
are normalized to the non-ionizing energy loss (NIEL) of 1 MeV neutrons (n$_{\rm{eq}}/$cm$^2$).}, 
the second and third layer to about $1.2\times10^{14}$ n$_{\rm{eq}}/$cm$^2$/yr and 
$0.6\times10^{14}$ n$_{\rm{eq}}/$cm$^2$/yr, respectively.
All components of the pixel system are specified to stay operational up to a
particle fluence of at least $6\times10^{14}$ n$_{\rm{eq}}/$cm$^2$. This
implies that parts of the detector will have to be replaced during the 
lifetime of the experiment. In case of a possible luminosity upgrade of the
LHC the particle fluence will be even higher.
For this reason it is necessary to test whether the detectors can be operated
at fluences above the ones specified above.

Particle irradiation affects the silicon sensor response in various ways.
It is common practice to divide radiation damage into surface and bulk damages.
The former includes all effects caused by io\-ni\-za\-tion in the dielectric covering
the structured sensor surface. Most important is the increase of charge in the silicon oxide
which saturates after some kGy to values of a few $10^{12}$ cm$^{-2}$. This changes
the electric field close to the collecting electrodes and may lead to electric breakdown. 
At higher particle fluences bulk damage, caused by the interaction of
the incident particles with the silicon lattice atoms, also becomes important.
Silicon atoms can be ejected from their lattice sites leaving
vacancies, and the recoil atoms can collide with other lattice atoms.
With large enough recoil energy other atoms are displaced, 
creating more vacancies and interstitials 
(also called {\it Frenkel pairs} or {\it primary defects}). 
The thermal energy of the crystal enables some of the simple defects
and defect clusters to migrate through the lattice and to react with other
defects or impurities (e.g. dopants such as phosphorous and boron or 
crystal impurities such as oxygen and carbon).
The formation of complex defects produces new energy states in the band gap, 
some of which can be electrically charged. 

There are mainly three macroscopic manifestations of the defects in reverse biased silicon 
detectors:
An increase of the leakage current, the modification of the space charge density
and of the drift field across the bulk, and the trapping of charge carriers.

The increase of leakage current leads to a higher power dissipation
and to a rise of the sensor temperature. Proper cooling is required to prevent
a destructive ``thermal runaway'' of the device. As the pixel current is
still very small the noise increase is not critical. However, the preamplifier 
DC-coupled to each pixel must be able to drain the excess of leakage current.
The formation and annealing of the crystal defects show a complex time 
behaviour which is strongly temperature dependent. 
It is advantageous to keep the sensors cooled to
about -10$^\circ$C not only during operation but also during shutdown 
periods to inhibit reverse annealing. In addition, it was
shown that radiation tolerance can be improved by enriching the starting
material with oxygen~\cite{Lindstrom:2001ww}.

Trapping of the mobile carriers from the leakage current produces a net positive space charge density near the p$^+$-backplane and a net negative space charge density near the n$^+$-implant.  Since positive space charge density corresponds to n-type doping and negative space charge corresponds to p-type doping, there are p-n junctions at both sides of the detector.  Consequently, the electric field profile across the bulk varies with an approximately
quadratic dependence having a minimum at the zero of the space charge density and maxima at both implants~\cite{Li:1992,Beattie:1998fw,Casse:1998hy,Eremin:2002wq, Verbitskaya:2003eg,Chiochia:2004qh,Chiochia:2005ag,Swartz:2005vp}.
The energy levels associated with the crystal defects are not only filled by
charge carriers from the generation-recombination current (leakage current) but
also by carriers produced by traversing particles. If these carriers are trapped
for a time longer than the signal collection time they do not contribute to
particle detection and are de facto lost. Trapping can be reduced by
collecting electrons instead of holes, because they have a higher mobility  
and are less prone to trapping. In addition, high drift fields and fast 
signal collection can be achieved by increasing the bias voltage.
Signal reduction caused by trapping presently limits the use of silicon
detectors in environments exposed to fluences higher than 10$^{15}$ n$_{\rm{eq}}/$cm$^2$.

The optimization of the CMS barrel pixel sensors was achieved through an
extensive comparison of several technological choices~\cite{Bolla:2001ra,Kaufmann:2001,
Bolla:2002my,Bolla:2003si,Dorokhov:2003if,Rohe:2004cm}. 
In particular, pixel sensors with p-spray and p-stop isolation of the collecting
electrodes were compared and the p-spray technique was chosen for the 
barrel design~\cite{Rohe:2004cm}. Therefore, only the results obtained with pixel 
sensors implementing a p-spray isolation are discussed hereafter. 
The purpose of the beam test program described in this paper was to investigate several
aspects of the sensors response:
\begin{itemize}
\item Prove that after a given irradiation fluence the bias voltage could be
adjusted to collect sufficient charge. In addition, due to the sparsified readout, sensor
regions with lower charge collection can lead to detection inefficiencies which may
be reduced by design optimization.
\item A good spatial resolution must be guaranteed during operation. 
In the CMS transverse plane most tracks are perpendicular to the sensors and 
charge is distributed among several pixels by the Lorentz deflection which
is affected both by the applied bias voltage and by irradiation.
Along the longitudinal direction larger impact angles produce clusters of several pixels and
the unhomogeneous charge collection caused by trapping produces 
asymmetric clusters. Thus, appropriate hit reconstruction techniques have to be developed to 
maximize precision.
\end{itemize}

This paper presents a detailed description of the sensors for the CMS barrel pixel
detector and their performance. Emphasis is put on highly irradiated sensors.
The paper is structured as follows: A detailed description of the pixel sensors layout
is given in Section~\ref{sec:sensor_description}. Both the final and prototype sensors used
for the tests are discussed. 
The samples used for the beam tests are described in Section~\ref{sec:sample_preparation}
and the electrical measurements are presented in Section~\ref{sec:electrical_measurements}.
The setup developed for the beam tests is described in Section~\ref{sec:experimental_setup} 
and the data reduction discussed in Section~\ref{sec:data_analysis}. The results of the beam test measurements
are presented is Section~\ref{sec:results}. Conclusions are presented in Section~\ref{sec:conclusions}.


%
%
\section{Sensor Description\label{sec:sensor_description}}

\subsection{Sensor dimensions and geometry\label{sec:geo}}

For the CMS barrel pixel detector two different sensor geometries, 
the so-called {\em full} and {\em half} modules are required. 
The full modules house $2\times 8$ readout chips. The half modules
have only $1\times 8$ readout chips. Each readout chip contains
$52\times 80$ pixels with dimensions of $150\times 100\,\mu$m$^2$ \cite{CMSTrackerTDR:1998}.

For the beam test measurements presented in this
paper a special readout chip with $22\times 32$ pixels 
of $125\times 125\,\mu$m$^2$ was used. Special sensors were designed to
match the chip dimensions. The pixel cell was scaled keeping the 
gap between the implants constant in order not to change the 
charge collection properties of the sensor. All other design 
features were kept identical to the final sensors. Therefore
it is not expected that the basic properties other than the 
spatial resolution differ significantly from the final barrel sensor design.

\subsection{Technological choices}

The sensors for the CMS pixel detector adopt the so
called ``n-in-n'' concept: Pixels consist of high
dose n-implants introduced into a highly resistive n-substrate.
The backside of the substrate is p-doped, therefore
the pn-junction is placed on the backside
of the sensor. As the junction must not extend to the diced
edge of the sensor a structured back side and, consequently,
a double-sided processing of the wafers is mandatory.
Additional precautions have to be taken to suppress the
electron accumulation layer shortening the  pixels on
the front side of the sensor. Both considerations lead
to a significant cost increase compared
to the single sided ``p-in-n'' sensors widely used for
strip detectors. 
However, this concept was chosen for the following reasons:
\begin{itemize}
\item It assures a high signal charge at moderated 
	bias voltages ($<600\,$V) 
	after high hadron fluences (see Sect.~\ref{sec:fluence_dep});
\item The high mobility of electrons leads to a larger 
	Lorentz angle, important to reach the required spatial resolution;
\item The need to structure the back side allows to implement a guard ring
	scheme keeping all sensor edges at ground potential.
\end{itemize}
Two inter-pixel isolation techniques were evaluated for the barrel sensors: 
Open p-stops \cite{bolla01} 
and moderated  p-spray \cite{mod-pat}. The moderated p-spray 
technique was eventually chosen because of the superior charge collection 
properties obtained with the prototype sensors \cite{ieee03}. 
The p-spray consists of a medium dose boron implantation 
``sprayed'' without mask over the whole wafer. 
Due to the topology on the wafer 
surface, the implant is ``moderated'' in the region of the
lateral pn-junction between pixel implant and the isolation,
as sketched in Fig.~\ref{fig:mod}. 
The implantation parameters are adjusted in a way that in the moderated region 
the boron dose matches as close as possible the expected saturation value of the 
oxide charge, reached after a few kGy of ionising radiation. Far from
the lateral pn-junction the boron dose reaching
the silicon can be roughly 2-3 times larger.
This improves the  high voltage stability of the unirradiated devices 
compared to a ``non-moderated'' p-spray \cite{rar96} while keeping the good
post-radiation behaviour.

As ground material n-doped FZ-silicon with $\langle 111\rangle$ orientation 
and a resistivity of $3.7\,\mathrm{k}\Omega$cm was used. All wafers for 
the production of the barrel sensors come from the same silicon ingot 
to provide the best possible homogeneity of all material parameters.
To improve the post radiation behaviour the wafers undergo 
a oxygen diffusion process as recommended by the ROSE 
collaboration \cite{rose} resulting in the so-called DOFZ-material.

\subsection{Mask layout}

\begin{figure}[hbt]
\centering\includegraphics[width=1\linewidth]{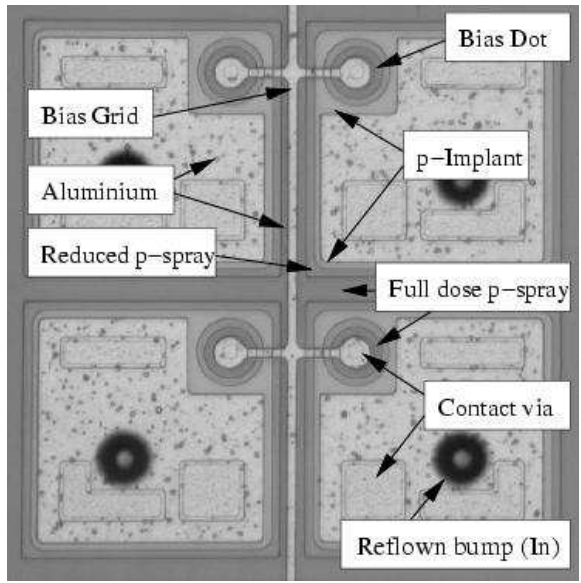}
\caption{Photo of four pixel cells. 
  \label{fig:pixellayout}}
\end{figure}
A photograph of four pixels is shown in Fig.~\ref{fig:pixellayout}.
The pixel size was in this case defined by the choice of the readout chip for
the test beam. While the CMS detector will be equipped with pixels
of $150\,\mu$m length along the beam pipe and $100\,\mu$m width in the polar direction,
the tested prototypes had dimensions of $125\times125\,\mu$m$^2$.

Most of the pixel area is covered with the collecting electrode formed by
the n-implant. The gap between the n-implants 
is kept rather small ($20\,\umu$m) to provide a homogeneous drift field for 
the signal charge. On the other hand, this small gap leads to a relatively high
capacitance of the order of 80-100\,fF per pixel which is still compatible
with noise requirements. In order to reduce electric field peaks the corners 
of the implant are rounded. 

The so-called {\em bias dot} is visible in one corner of each pixel (Fig.~\ref{fig:pixellayout}). 
It consists of a circular n-implant isolated from the pixel implant by a small gap.
It is connected through a contact via to a circular metal pad. 
These pads are routed to a metal line running along every second pixel column. 
They form a bias grid which provides a high resistive punch-through connection to all pixels. 
The grid allows on-wafer current-voltage (IV) measurements prior to bump bonding which are
performed to detect faulty sensors. In the assembled module
the pixels are grounded by the bump bond connection to the preamplifier 
(DC-coupling) and the grid is used to bias accidentally unconnected pixels.

\begin{figure}[hbt]
\centering\includegraphics[width=1\linewidth]{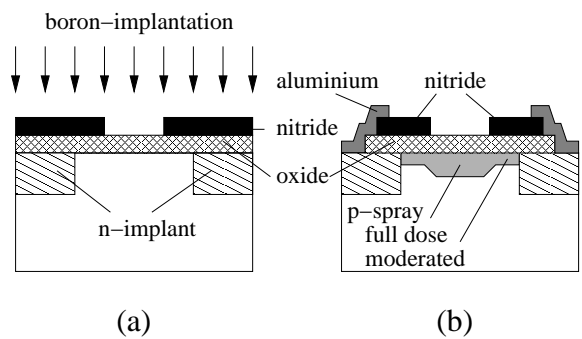}
\caption{Moderated p-spray isolation: (a) topology on the wafer surface
	during implantation, (b) resulting doping profile 
	in the fully processed device.\label{fig:mod}}
\end{figure}

The dark ``frame'' around the pixel implants visible in Fig.~\ref{fig:mod} 
indicates the opening in the nitride covering the thermal oxide.
In this region the p-spray dose reaches its maximum. The boron dose 
is reduced close to the lateral pn-junction between the pixel implant and the p-sprayed
inter-pixel region. To increase the punch through voltage 
the full p-spray dose is applied between the bias dot and the pixel implant.
This is possible as the gap is small compared to the gap between the pixels and both gaps are 
electrically isolated from each other another.

The pixel metalization is slightly smaller than the pixel implant. An
overhanging field plate could not be implemented as the metal connection 
of the bias grid had to cross the edge of the pixel implant.
The metalisation is connected to the underlying implant via several contact holes.
Since too large openings might reduce the production yield, their total area was kept small.

The sensor shown in Fig.~\ref{fig:pixellayout} has undergone the bump
deposition process including the reflow of the bumps. The bumps are Indium
droplets approximately 25 $\mu$m in diameter. Due to their finite size 
they are out of the focus of the microscope.

The active area of the module is surrounded by a $180\,\mu$m wide 
n-implanted ring to which all metal lines from the bias grid are connected.
This bias ring can be contacted either by large wire bond pads 
in the sensor corner or via a special bump pad on the readout chip, although
this is currently not foreseen. The region outside this structure spans a large
area n-implant up to the scribe line. It will be connected to the analogue ground
of the readout chip via a second special bump pad. This connection ensures
that the whole sensor edge is kept at ground potential and in addition drains
the current from the guard ring region. In the test samples investigated during 
this study this connection was open and the outer region
was floating. The edge current was therefore not drained but equally distributed
over all pixels via the bias grid. However, the noise increase 
produced by the extra current flowing into the preamplifier was negligible.

\begin{figure}[hbt]
  \centering\includegraphics[width=1\linewidth]{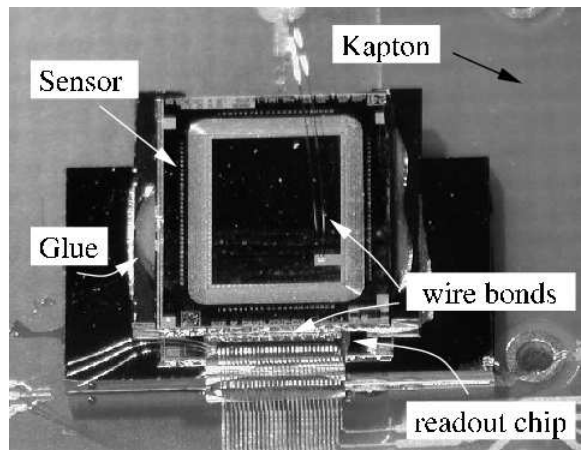}
  \caption{Photograph of a sensor attached to the readout chip.
	The glue guaranteeing the sample stiffness and the 
	wire bond connections are visible.
	\label{fig:mounted_sample}}
\end{figure}

The back side of the sensor consists of a large p-implant forming the
junction that depletes the sensor. This p-implant overlaps the 
sensitive region by $100\,\umu$m to provide a homogeneous drift field 
also in the edge pixels. The metalization is open in almost the
whole sensitive area apart from the wire bond pad for the 
bias voltage, as visible in Fig.~\ref{fig:mounted_sample}. This 
allows charge injection by a laser.

The sensor is surrounded by a multiguard ring structure, shown in
Fig.~\ref{fig:mounted_sample}, that guarantees 
a gradual potential drop to the scribe line and prevents the lateral extension 
of the space charge region. It consists on 16 p-implanted guard rings with 
a distance increasing from $10~\mu$m in the innermost ring to $50~\mu$m in the
outermost. The aluminisation of the rings is overhanging in both directions.
A small metal overlap directed toward the scribe line reduces the electric field
at the implant edge. A larger overlap toward the device centre
increases the voltage drop between rings \cite{bischoff,avset}.

%
%
\section{Sample preparation\label{sec:sample_preparation}}

The single chip sensors used in the test beam measurements
(Section~\ref{sec:results}) were produced 
in 2002 by CiS, Erfurt, Germany. The deposition of 
the under bump metalisation~(UBM) and the indium bumps on wafer was performed at PSI. 
Then the wafers were diced and the bumps reflown. Some of the sensors 
were bump bonded to the readout chips and glued to support cards. 

The treatment of irradiated sensors is more complex. 
Sensors were irradiated before bump bonding. This was required to study the
effects of irradiation on sensors only. In addition, the readout chip 
used for this study was not sufficiently radiation hard.
The bump deposition and reflow involve temperatures above 
$120^\circ\,$C. To prevent reverse annealing these steps had to be 
performed before irradiation. 

\begin{figure}[hbt]
\centering\includegraphics[width=1\linewidth]{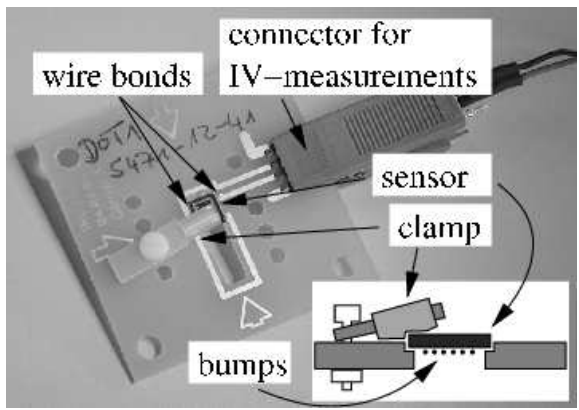}
\caption{Photograph of a sensor carrier for irradiation. The sensor is
	held by a clamp and contacted with wire bonds. After irradiation
	and IV-measurements the sensor can be removed and attached to the
	readout chip.\label{fig:carrier}}
\end{figure}

For irradiation the diced sensors were mounted on 
the PCBs shown in Fig.~\ref{fig:carrier}.
These $50\times 50\,$mm$^2$ PCBs can safely be transported
in slide holders, thus the fragile bumps are protected. The
wire bond connection to the sensor's back side easily allows 
IV-measurements.

The irradiation has been carried out in 2003 and 2004 at the CERN-PS 
using 21-24 GeV protons. The delivered proton fluences scaled to 1\,MeV
neutrons by the hardness factor 0.62 \cite{rose} were between 
$4.7\times 10^{13}$ and $2.6\times 10^{15}\,\Neq$. The irradiation was 
performed without cooling and no bias applied. After irradiation
the samples were stored in a freezer at $-18^\circ\,$C
and warmed up to room temperature only for handling and transport.

For mating the sensor to the readout chip a special procedure without
heat treatment was developed. The mechanical strength of the
bump bond connections usually obtained after a second reflow was
achieved by the application of glue (see Fig.~\ref{fig:mounted_sample}).
The assembly was then attached to a Kapton card an the electrical connections
were established by wire bonds. The Kapton cards can be inserted 
into the cooled rotating stage of the beam telescope (see Fig.~\ref{FIG:telescope} below).

After final assembly the samples underwent annealing of three days at $30^\circ$\ C. 
This was done to adjust the depletion voltage close to its minimum according to 
the Hamburg model for DOFZ material \cite{rose}.
%
%
\section{Electrical measurements\label{sec:electrical_measurements}}

\begin{figure*}[hbt]
\centering\includegraphics[width=1\textwidth]{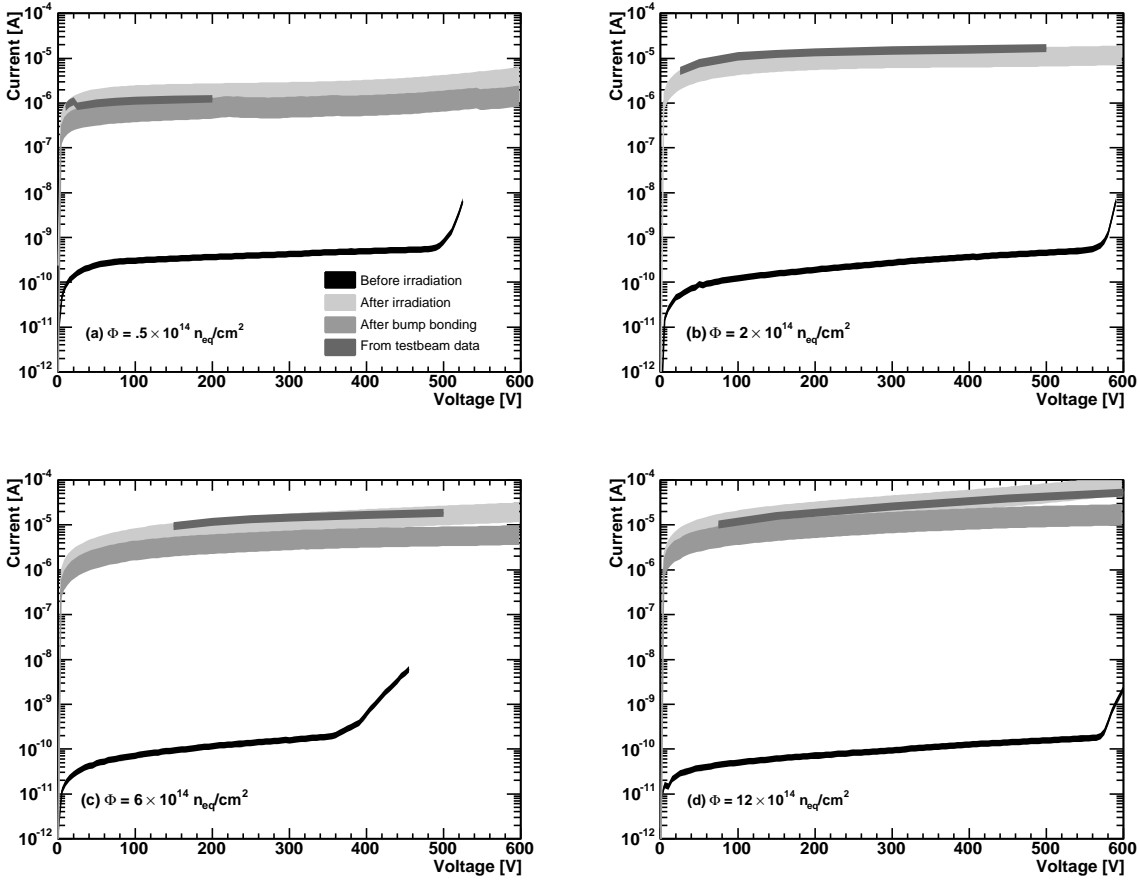}
\caption{IV curves of the four detectors used for the measurements discussed in Section~\ref{sec:results}. 
  The current is normalized to $-10^\circ$\ C. 
  The shaded bands show the uncertainty due the temperature measurement.
  \label{fig:iv}}
\end{figure*}

To assure that the samples were not mechanically damaged during
the long and complex preparation, IV-characteristics were
measured before each important assembly step:
\begin{enumerate}
\item Shortly before irradiation, after mounting and
  wire bonding the sample to the support structure.
  \item A few days after irradiation, when
  the sample was still mounted on the support card.
\item After bump bonding and mounting on the Kapton card.
\item During the beam test.
\end{enumerate}
Sensors with obvious damage leading to low breakdown
voltage or extremely high leakage current were excluded from further
assembly. The IV-curves of the sensors used for the measurements
discussed in Section~\ref{sec:results} are shown in Fig.~\ref{fig:iv}.
The current is normalized to $-10^\circ$\,C using the temperature 
dependence of the volume generation current:
\begin{equation}
I\propto T^2 \e^{-E_g/2kT},
\end{equation}
where $T$ is the absolute temperature, $E_g$ the band gap in
silicon and $k$ the Boltzmann constant. 
The temperature was measured using a platinum thermistor (PT100) located 
close to the sensor. The error bands visible
in Fig.~\ref{fig:iv} are due to the uncertainty on the temperature,
which was estimated to $\pm 2^\circ$\ C for the measurement of
the unirradiated sensor and the data extracted from the test beam slow
control system and $\pm 5^\circ$\ C for the other measurements. 
The temperature difference between the pixel sensor and the thermistor 
for bump bonded devices was estimated by comparing the sensor leakage 
current with and without power applied to the frontend chip.

The expected leakage current~\cite{moll} is in the lower part of the error bands.
A factor of about two can be explained by the extremely 
large edge area of the small sensors (the guard current is not drained separately).
The high mechanical stress caused by the special bump bonding 
process involving glue application may also leads to a current increase.

More important than the level of the leakage current is the 
shape of the IV-curve. While the unirradiated sensors show a 
breakdown voltage between 400 V and 600\, V, no 
breakdown is observed after irradiation.
This proves the high radiation tolerance of the moderated p-spray technique
used for these devices.


%
%
%
%
%
\section{Beam test setup\label{sec:experimental_setup}}

The measurements were performed in the H2 beam line of the CERN SPS
in 2003-04 using 150-225 GeV pions. The test setup was placed
in an open-geometry 3~T Helmoltz magnet pro\-du\-cing a magnetic
field parallel or perpendicular to the beam.

\subsection{Beam telescope}

\begin{figure}[hbt]
  \begin{center}
    \epsfig{file=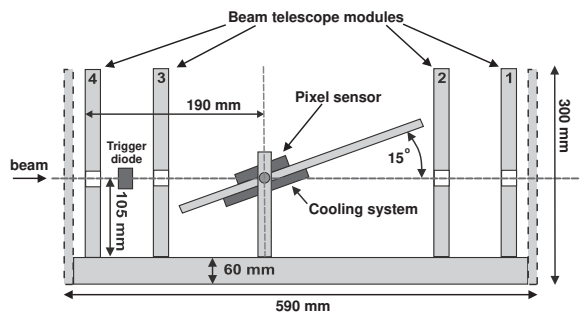,width=\linewidth}
    \caption{Side view of the beam test apparatus consisting of four horizontal and four 
      vertical planes of silicon strip detectors and a rotating stage for the pixel detector.}
    \label{FIG:telescope} 
  \end{center}
\end{figure}

A silicon reference telescope~\cite{Amsler:2002ta} was used to allow a precise determination of the particle hit coordinates in the pixel detector (see Figure~\ref{FIG:telescope}). The telescope modules and the hybrid pixel detector were mounted on a common frame. The beam telescope consisted of four modules, each including two orthogonal 300 $\mu$m thick single-sided silicon strip detectors to measure the horizontal and vertical impact coordinates. The sensors have a strip pitch of 25 $\mu$m, a readout pitch of 50 $\mu$m and a total sensitive surface of $32\times30$ mm$^2$. The intrinsic position resolution of the beam telescope is about 1~$\mu$m~\cite{Amsler:2002ta}.

The pixel hybrid detector and the readout card are mounted between the second and third 
telescope module on a cooled rotating stage. The rotation is around the horizontal axis 
and a precision of 0.2$^\circ$ is achieved.
A trigger signal is generated by a silicon PIN diode placed between the first and second telescope modules. 
The horizontal and vertical positions of the diode can be adjusted with a precision of about 100 $\mu$m.

\subsection{Pixel sensors assembly and cooling}

The hybrid pixel sensor is enclosed in a thermo--isolated plastic box
flushed with dry nitrogen to avoid water condensation. The readout chip  is bump-bonded
to the pixel sensor and glued to a printed circuit board (PCB). The pixel readout card is
connected to the PCB via a short flat cable.

The power consumption of the pixel readout chip and the readout card is about 20 W. 
This heat has to be efficiently removed by the cooling system.
Therefore, two high performance Peltier elements attached to a liquid cooled heat sink are
used to chill the sensor.  The hot ceramic side of the Peltier elements is in direct 
contact with the coolant liquid, which is circulated by an external Lauda WKL600 chiller.

The temperature is measured with a PT1000 platinum resistor placed on the PCB
and stabilized by regulating the voltage applied to the Peltier elements. 
Heat conducting paste is used to improve the thermal contact between the cold aluminum plates 
and the PCB. The sensor resistance is measured with a Keithley 2000 multimeter 
and the temperature can be stabilized down to -35$^\circ$~C. 

\subsection{Sensors readout and data acquisition}

\begin{figure*}[hbt]
\begin{center} 
  \epsfig{file=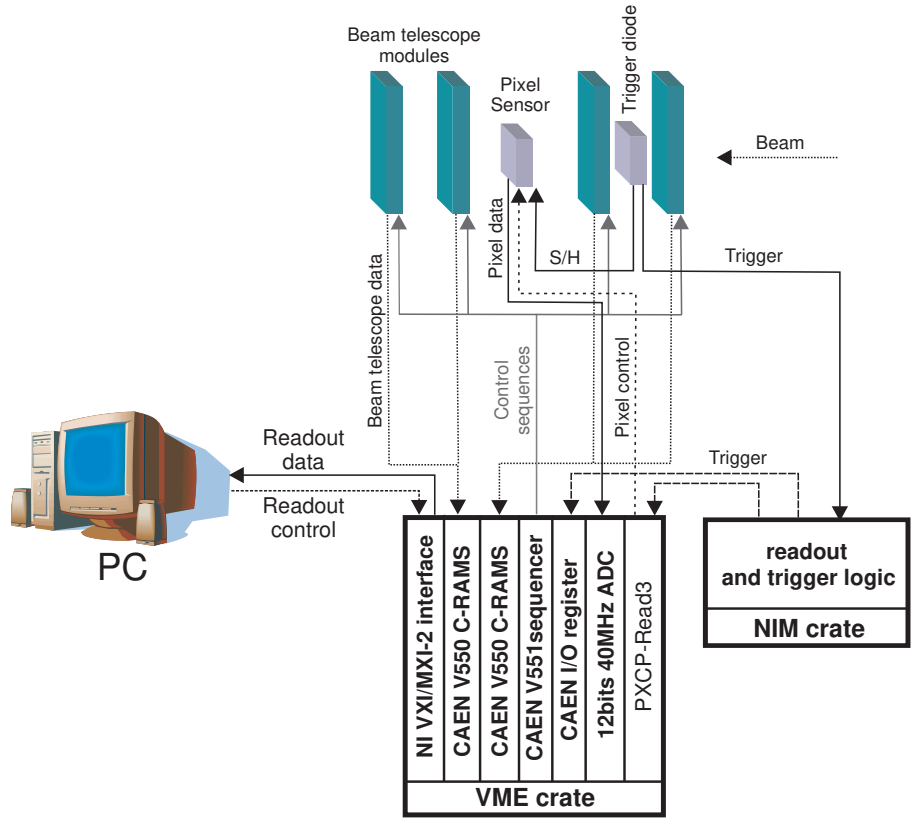,width=12cm,clip=,angle=0,silent=} 
\end{center}
\caption{Schematic diagram of the setup readout.}
\label{testbeam_sch.eps}
\end{figure*}

A schematic view of the readout is shown in Fig.~\ref{testbeam_sch.eps}.
The analog signals of the 22$\times$32 pixels are read out 
by the PSI30/AC30 readout chip~\cite{Meer:2000}.
Each pixel cell is read out by a preamplifier and a shaper  
with a peaking time adjusted to 45 ns.
The readout is enabled when the particle crosses the PIN diode area and
a trigger pulse is sent to the sample--and--hold (S/H) input 
of the readout chip. The active level of the S/H signal disconnects 
a capacitor attached to the shaper output and the signal amplitude 
is stored in the capacitor.
In addition, the signal discrimination is disabled to allow
the readout of all cells.
The analog amplitudes are digitized by a 12 bits 40 MHz ADC.

The trigger signal is also used to start the readout of the beam
telescope modules. The charge collected by the strip sensors is amplified
and  integrated in a time window of about 2 $\mu$s. 
The signals are then digitized by two CAEN V550 units controlled by a 
CAEN V551 sequencer.

The data acquisition (DAQ) software is written in LabView and LabWindows CVI\footnote{LabView 
and LabWindows/CVI are products of National Instruments.} and runs on a PC.
It controls the VME crate via a National Instruments PCI interface, 
the power supplies for the  pixel sensor and the Peltier elements, 
and the multimeter for the temperature measurements.

The pixel settings and trigger types can be changed. 
The telescope, pixel signal amplitudes and pedestal values are stored in a
data file. Each event is marked with a time stamp and the values of the pixel 
bias voltage, dark current and temperature are recorded.
%
%
\section{Data analysis\label{sec:data_analysis}}

\subsection{Event selection\label{sec:event_selection}}

The channel noise and pedestal levels are measured with dedicated 
random trigger runs taken during the spill gaps. 
The signal amplitudes of each pixel cell is reconstructed with 
the following procedure~\cite{Dorokhov:2005}: 
\begin{enumerate}
\item
  Pixel pedestals stored in the data file are subtracted from 
  the amplitudes cell by cell.
\item
  The amplitudes of six unconnected pixel rows 
  are averaged and subtracted to reduce common mode fluctuation.
\item
  Hit pixels above threshold (which depends on the irradiation fluence),
  are further analyzed. The threshold is optimized for each sample and ranges
  between 35 -- 50 ADC counts (corresponding to 1000 -- 1500 electrons).
\item
  If the hit frequency is below 0.1 times or above 10 times the expected one, 
  the pixel is marked as noisy or dead and excluded from the amplitude reconstruction.
\item
  Improved pedestals and common mode values derived from the good pixels 
  with amplitudes below threshold, are calculated and subtracted from the 
  pixel amplitudes.
\end{enumerate}
Steps (iii), (iv) and (v) are repeated for four iterations.
The signal amplitudes of the telescope strip sensors are reconstructed 
and corrected in a similar way.

Events used for the
alignment of the pixel detector with the beam telescope are selected.
The selection procedure discards the events with multiple tracks 
and tracks which do not fit to straight lines. 

\subsection{Alignment of the pixel sensor with the beam telescope\label{sec:alignment}}

In each beam telescope plane the position of the impact point is reconstructed 
with a center of gravity algorithm for clusters of two adjacent strips.
The main strip in the cluster must have a signal between 30 and 350 ADC counts and the
second strip is the next neighbour with the highest amplitude.
The hit position is then corrected using the $\eta$ algorithm described 
in~\cite{Belau:1983eh}.

\begin{figure}[hbt]
  \begin{center}
    \epsfig{file=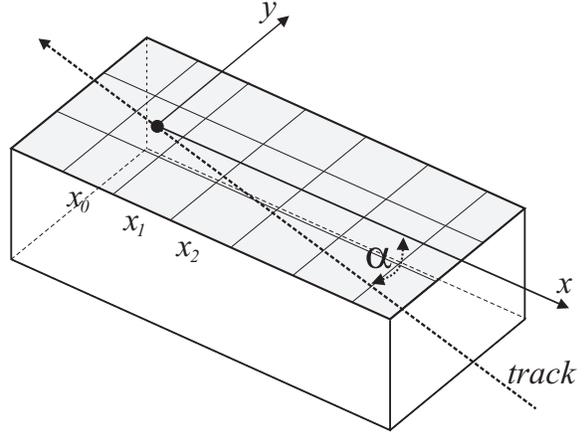,width=\linewidth}
    \caption{Reconstruction of the impact position in the pixel sensor.}
    \label{FIG:15deg_position_reconstruction} 
  \end{center}
\end{figure}

The determination of the particle impact position in the pixel sensor
depends on the track incident angle $\alpha$ with respect to the pixel plane
(see Fig.~\ref{FIG:15deg_position_reconstruction}).
For $\alpha>60^\circ$ the hit position is measured using a center of gravity
algorithm. The pixel with the highest signal above threshold is located
and the surrounding eight pixels are included in the cluster. The cluster
is projected along the two coordinates $x$ and $y$. The weighted averages of
the hit coordinates $x$ and $y$ are calculated.

For small angles, e.g. $\alpha=15^\circ$, the cluster size along $x$ is in
general larger than three pixels and a different approach is used.
The pixel corresponding to the track exit point is defined by the first
pixel above threshold along the $x$ axis and with the highest signal along
the $y$ axis. A $3\times3$ cluster is constructed by including the following two
pixels along the $x$ axis, which are required to be over threshold, and the 
corresponding three pixels in the upper and lower rows along the $y$ axis.
The $y$ coordinate of the cluster is determined with the center of gravity
of the signal collected in the three $x$-rows. The $x$ coordinate of the 
track exit point is determined using the signal in the three cluster $y$-columns.
We define the signals $A_0,A_1,A_2$ and the $x$ coordinates $x_0,x_1,x_2$ 
of the three pixel columns, respectively. The signal sums $S_i=\sum_{j=0}^i A_j$ with $i=(0,1,2)$
vs. their $x$ coordinates are fitted with a straight line.
The intercept of this line with the $x$ axis gives the coordinate of the
track exit point.

The alignment is performed by minimizing the residuals in the following system of equations
\begin{eqnarray}
  \sum_{i=1}^{8} \left[ p_{ij}  C_{xi} \right] + C_{x0}  - x_{j} & = & R_{xj}, \\
  \sum_{i=1}^{8} \left[ p_{ij}  C_{yi} \right] + C_{y0}  - y_{j} & = & R_{yj}, 
\end{eqnarray}
where $p_{ij}$ is the position in the $i$-th beam telescope plane\footnote{The indexes $i=[1,...,4]$ correspond
  to the position in the $x$ planes, while $i=[5,...,8]$  to the positions in $y$ telescope planes.} 
for the $j$-th event, $x_{j}$ and $y_{j}$ are the hit coordinates in the pixel coordinate system,
$R_{xj}$ and $R_{yj}$ are the residuals in the $x$ and $y$ coordinates,
respectively, $C_{xi}$ and $C_{yi}$ are the fit parameters for the $x$ and $y$ coordinates,
respectively.

\subsection{Gain calibration\label{sec:gain_calibration}}

\begin{figure}[hbt]
  \begin{center}
    \mbox{
      \subfigure[]{\scalebox{0.90}{
	  \epsfig{file=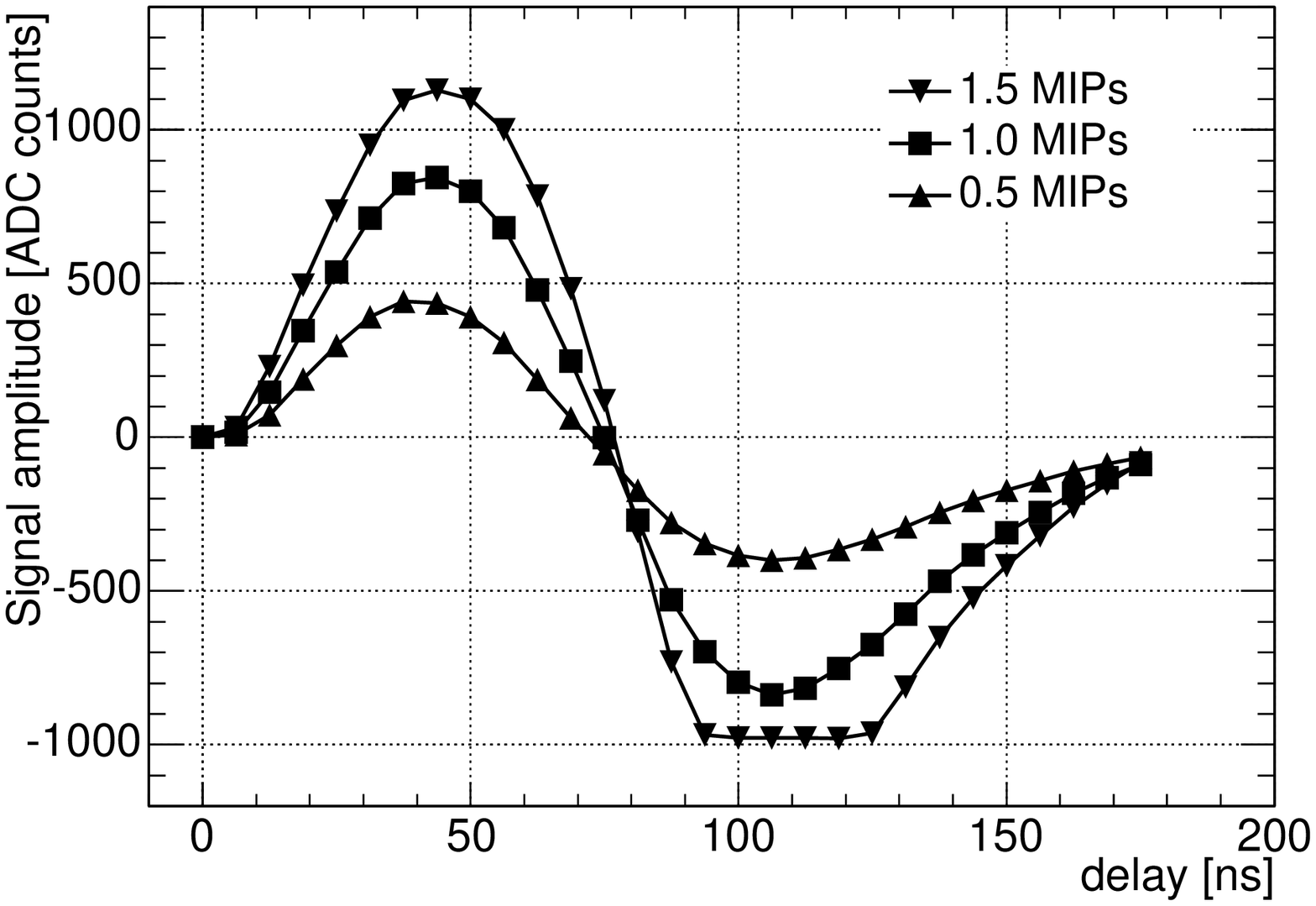,width=\linewidth}
	  \label{FIG:delay_scan}
      }}
    }
    \mbox{
      \subfigure[]{\scalebox{0.90}{
	  \epsfig{file=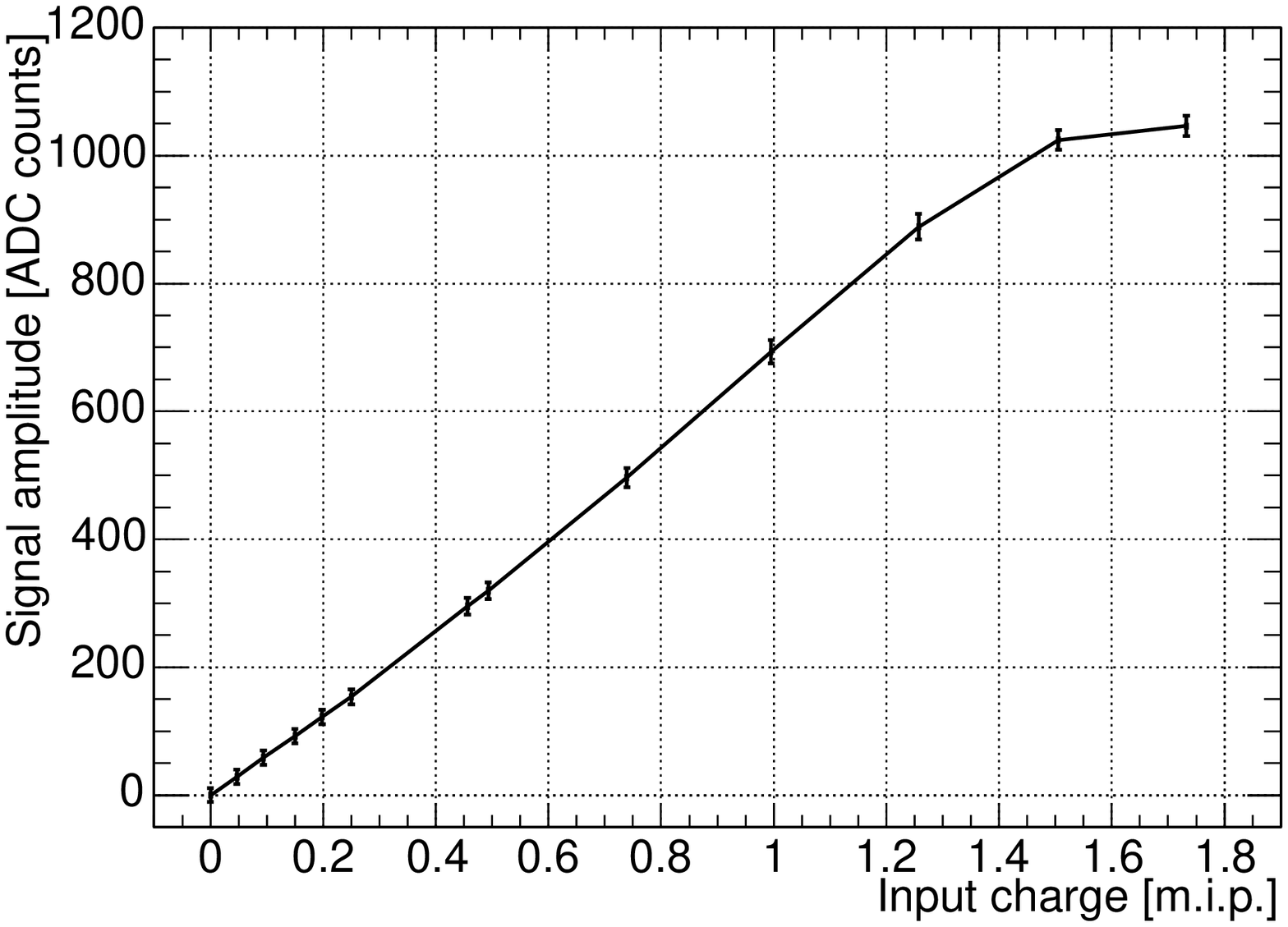,width=\linewidth}
	  \label{FIG:gain_calibration}
      }}
    }
    \caption{(a) Average signal amplitude as function of the time interval between the
calibration and trigger pulses for different values of the injected charge. 
(b) Average signal amplitude as function of injected charge. The delay is set to
55 nsec.} 
  \end{center}
\end{figure}

A calibration of the readout chain can be performed by injecting a known charge
at the input of the chip preamplifier via a 1.7 pF capacitor. The average signal
as a function of the time interval between the calibration and the trigger pulses
is show in Fig.~\ref{FIG:delay_scan}. The average signal measured with straight minimum
ionizing tracks and an unirradiated sensor is 710 ADC counts. This implies that the sampling
time of our measurements is about 55 ns. 
The gain calibration is performed by setting the delay to this value and measuring
the average signal amplitude as function of the charge injected in the preamplifier
(see Fig.~\ref{FIG:gain_calibration}). The charge released by a minimum ionizing particle (m.i.p.)
is 22400 electrons for a silicon thickness of 285 $\mu$m. 
The response of the frontend electronics is fairly linear up
to about 1.2 m.i.p. and saturates for higher values of the injected charge. 
Figure~\ref{FIG:gain_calibration} is used hereafter to translate the ADC
values into charge in electrons. More details about the gain calibration
can be found in~\cite{Dorokhov:2005}.
%
%
\section{Results~\label{sec:results}}

\subsection{Signal-over-noise ratio~\label{sec:sn_ratio}}
Signal, noise and their ratio measured after dif\-fe\-rent irradiation fluences and bias voltages
are summarized in Table~\ref{tab:sn_ratio}. The signal
was defined as the average signal recorded by the hit pixel with tracks perpendicular
to the sensor plane, while noise was the r.m.s 
of a Gaussian fit applied to the signal distribution recorded between spill gaps.
The measurements were performed with tracks perpendicular to the sensor surface without
magnetic field. As discussed above, the bias voltage must be increased after irradiation
in order to collect charge from the full sensor thickness. The optimal bias voltage
at each fluence was set to the minimum value at which charge collection saturates,
as discussed in Section~\ref{sec:fluence_dep}.
\begin{table*}[!Hhtb]
  \begin{center}
   \begin{tabular}{ccccc}
    \hline
   Fluence                        & Bias voltage & Signal       &  Noise        & S/N ratio \\
   $[\rm{n}_{\rm{eq}}/\rm{cm}^2$] & [V]          & [ADC counts] &  [ADC counts] &           \\
   \hline
   0                   & 150 & 727.7 & 11.1 & 65.2 \\
   $0.6\times10^{14}$  & 150 & 633.2 & 8.9  & 70.9 \\
   $2\times10^{14}$    & 200 & 654.0 & 10.9 & 59.8 \\
   $6\times10^{14}$    & 400 & 475.6 & 9.9  & 48.2 \\
   $12\times10^{14}$   & 600 & 430.5 & 22.8 & 18.9 \\
   $26\times10^{14}$   & 600 & 198.0 & 19.5 & 10.1 \\
   \hline
   \end{tabular}
   \caption{Signal, noise, and S/N ratio for different irradiation fluences and bias voltages.~\label{tab:sn_ratio}}
\end{center}
\end{table*}

The measured signal over noise ratio is high even after high fluences. 
At a fluence of $6\times10^{14}$ $\rm{n}_{\rm{eq}}/\rm{cm}^2$, 
roughly corresponding to the first four years of LHC operation of the innermost layer, 
it exceeds 40 which corresponds to a charge of 15500 electrons, roughly
65\% of the value measured with the unirradiated sensor.
By doubling the irradiation fluence the signal drops to 59\% of the unirradiated
sensor signal when the bias voltage is increased to 600~V. The noise becomes 
quite high and the signal over noise ratio decreases to about 20.
The high noise is partly caused by the absence of leakage current compensation
in the readout chip. 
Nevertheless, as we show in Section~\ref{sec:hit_efficiency}, it is still
possible to operate the sensor with a charge threshold of 2000 electrons maintaining 
the hit detection efficiency above 97\%. \\
The most irradiated sample is a factor of four above specifications and almost
in the range of irradiation fluences expected after LHC upgrade. 
The average signal is then about 6300 electrons. For efficient
particle detection the threshold has to be set below 3000 electrons and
the noise level kept below 600 electrons. This might be possible with
a readout chip implementing a well--suited leakage current compensation.

\subsection{Position dependence of charge collection~\label{sec:position_dep}}
The structured surface of the sensor can lead to spatial variations in charge collection.
We have therefore investigated the charge collection 
with tracks perpendicular to the sensor plane without magnetic field. 

\begin{figure*}[hbt]
  \begin{center}
    \mbox{
      \subfigure[]{\scalebox{0.40}{
          \epsfig{file=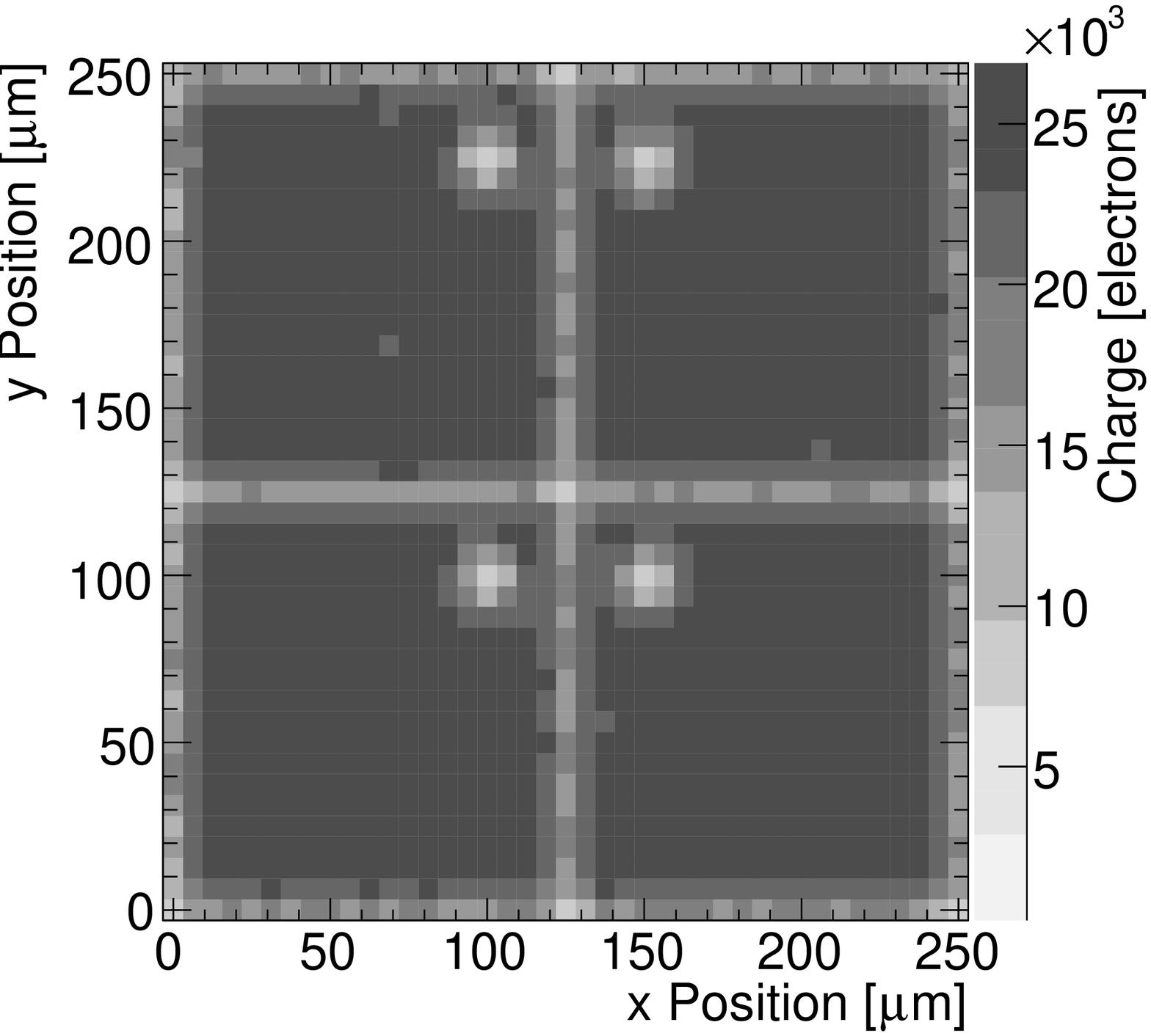,width=\linewidth}
          \label{fig:ch_vs_pos_a}
      }}
      \subfigure[]{\scalebox{0.40}{
          \epsfig{file=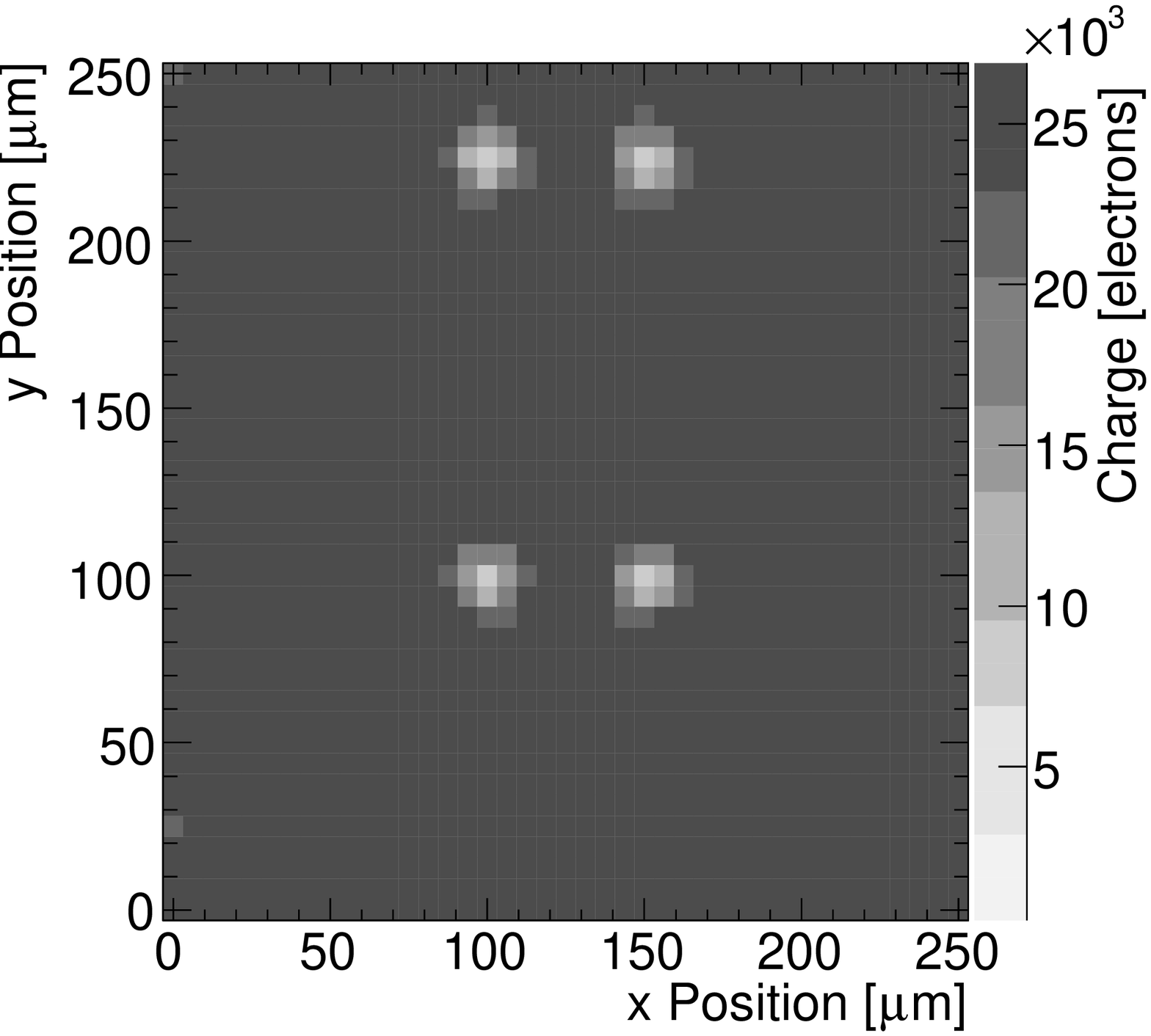,width=\linewidth}
          \label{fig:ch_vs_pos_b}
      }} 
    }
    \mbox{
      \subfigure[]{\scalebox{0.40}{
          \epsfig{file=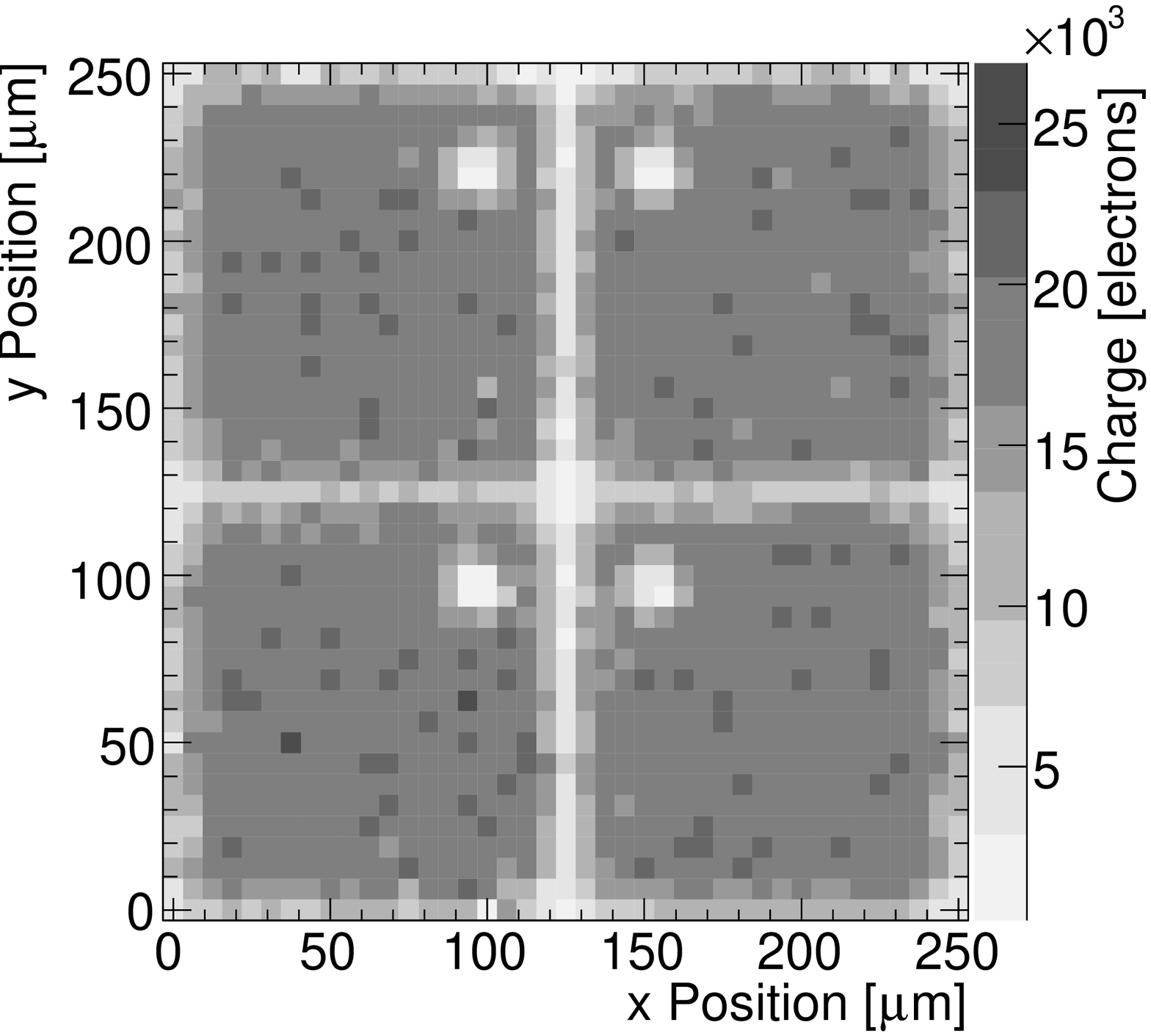,width=\linewidth}
          \label{fig:ch_vs_pos_c}
      }}
      \subfigure[]{\scalebox{0.40}{
          \epsfig{file=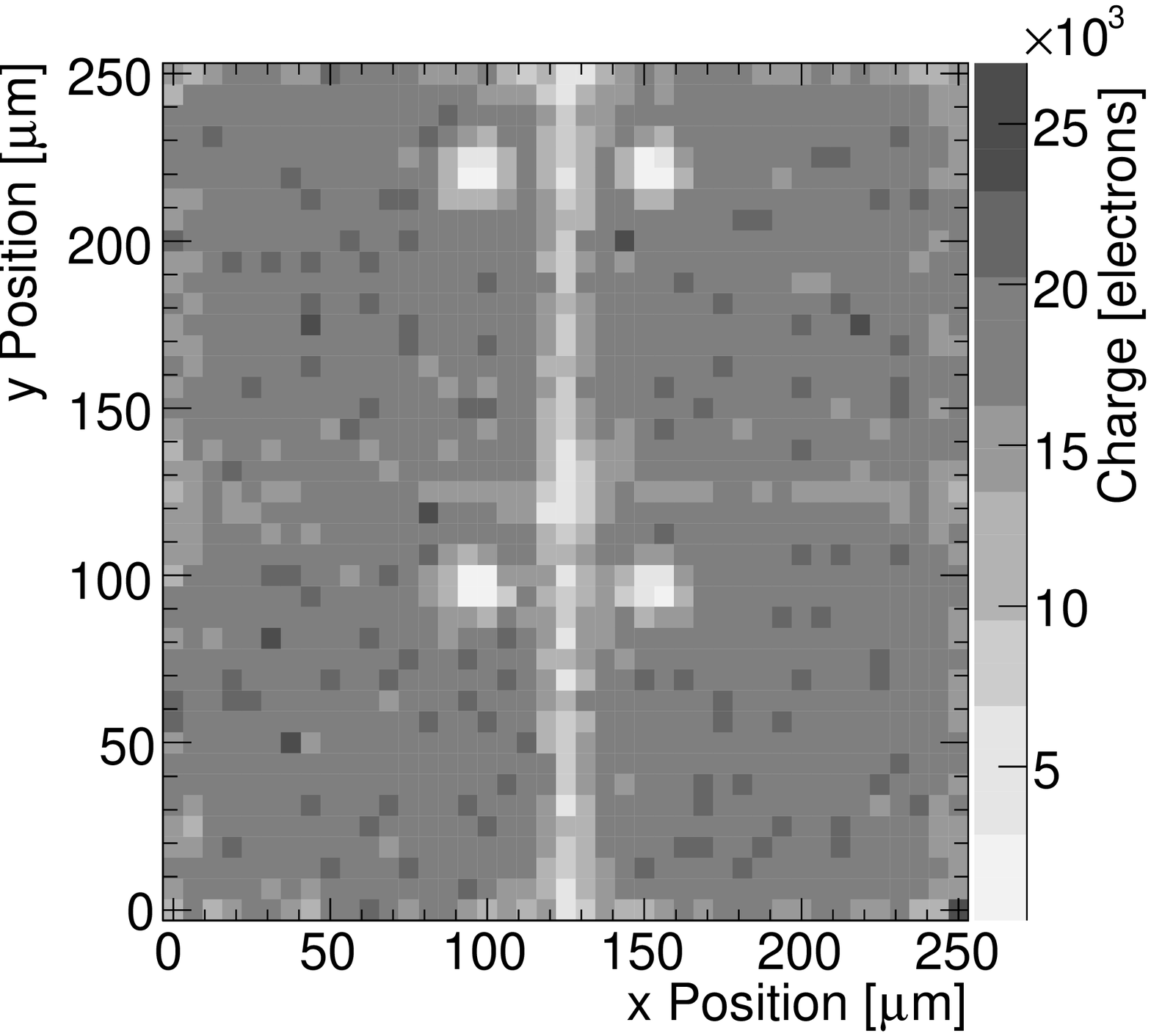,width=\linewidth}
          \label{fig:ch_vs_pos_d}
      }}
    }
    \caption{Average charge collected in the hit pixel ((a) and (c)) and in a $3\times3$
      cluster around the hit pixel ((b) and (d)) as function of the hit position. Each
      figure shows a pixel matrix of size $2\times2$. The charge
      is measured with an unirradiated sensor with bias voltage set to 150~V ((a) and (b))
    and with a sensor irradiated to $6\times 10^{14}$ n$_{eq}/$cm$^2$ with sensor
    bias set to 450~V ((c) and (d)).\label{fig:ch_vs_pos}}
  \end{center}
\end{figure*}

Figure~\ref{fig:ch_vs_pos} shows the average collected charge as a function of the
hit position predicted by the beam telescope for an unirradiated sensor operated
at 150~V (Fig.~\ref{fig:ch_vs_pos_a} and (b)) and a sensor irradiated to a fluence
of $6\times 10^{14}$ n$_{eq}/$cm$^2$ and operated at 450~V (Fig.~\ref{fig:ch_vs_pos_c} and (d)).
Shown are four pixel cells with the metal line along the vertical axis at 
the $x=125$ $\mu$m position. 
The left column (Fig.~\ref{fig:ch_vs_pos_a} and (c)) shows the charge collected
in the hit pixel while the right column (Fig.~\ref{fig:ch_vs_pos_b} and (d))
shows the charge in a $3\times3$ cluster around the hit pixel.

In the unirradiated sensor the charge is uniformly collected within the pixel implant 
with the exception of the punch-through biasing structure where the signal is reduced by up to 50\%.
The latter is located too far from the pixel edge so that the 
charge deposited in this position is not shared with other pixels.
The effect of the punch-through structure on the overall sensor performance 
is very limited as it represents only 2--3\% of the total surface.

In the region between pixel implants the charge summed over a $3\times3$ cluster
is not noticeably affected (see Fig.~\ref{fig:ch_vs_pos_b}). Charge sharing is restricted 
to a region of about 10 $\mu$m around the pixel edge, where the signal 
in the hit pixel decreases. It will be enhanced by the external 4 T 
magnetic field of CMS and also for non perpendicular tracks. 

After irradiation the charge collected in the hit pixel decreases by about 30\%
due to trapping of the charge carriers and the charge losses in the punch-through
structures become more prominent. The sum of the charge collected in the
hit pixel and in the surrounding neighbours is shown in Fig.~\ref{fig:ch_vs_pos_d}.
Charge sharing is uniform after irradiation along all pixel borders except
for the region around the metal line. 
Similar effects were observed with other hybrid pixel detectors implementing
a biasing grid~\cite{Lari:2002wb} and might be due to the capacitive coupling of the metal line to
the underlying n-bulk after irradiation.

\subsection{Hit detection efficiency~\label{sec:hit_efficiency}}

\begin{figure}[hbt]
  \begin{center}
    \epsfig{file=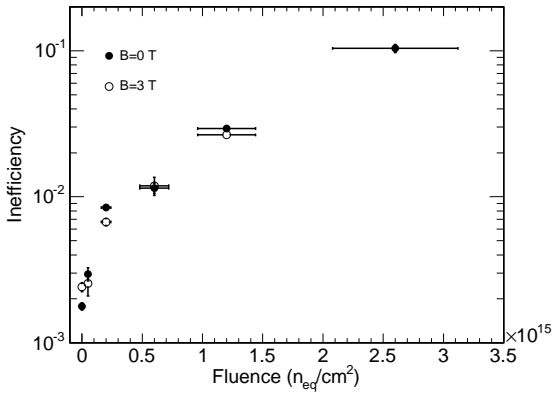,width=\linewidth}
    \caption{Hit detection inefficiency as function of irradiation fluence with
      (full dots) and without (empty dots) magnetic field. 
      The pixel threshold is set to 2000 electrons.}
    \label{fig:eff_vs_fluence} 
  \end{center}
\end{figure}

The sensor regions with reduced charge collection can be potential sources
of detection inefficiency when a threshold is applied.
The fraction of undetected hits, or detector inefficiency, was measured with tracks
perpendicular to the sensor plane.
The hit was counted as detected if the pixel determined by the beam telescope 
collected charge above a threshold of 2000 electrons. This value corresponds to the
foreseen threshold in the CMS experiment. Pixels cells with high
noise or with faulty bump-bond connections were excluded from the analysis.
Readout related inefficiencies were also excluded.
Figure~\ref{fig:eff_vs_fluence} shows the hit detection inefficiency as a function of
irradiation fluence with and without magnetic field. 
For each fluence the bias voltage was set to the values given in Table~\ref{tab:sn_ratio}.
The particle detection inefficiency is well below 1\%
for unirradiated sensors but it increases with irradiation fluence. 
The inefficiency depends only weakly on the applied threshold.  
After a fluence of $6\times10^{14}$ n$_{\rm{eq}}$/cm$^2$ the inefficiency 
is still below 2\% and acceptable for the operation within CMS.
For the sensors irradiated to fluences lower than $2\times10^{14}$ n$_{\rm{eq}}$/cm$^2$
the undetected hits are uniformly distributed within the pixel cell~\cite{Rohe:2004}. At higher
irradiation fluences the undetected hits are located mostly in the punch-through
structure and along the metal line. This is due to the reduced charge
collection in these regions, as discussed in Section~\ref{sec:position_dep}.

By applying a 3~T magnetic field parallel to the horizontal axis of Fig.~\ref{fig:ch_vs_pos}
charge carriers are deflected from the electric field lines and charge is spread over a region 
of about 50-130 $\mu$m, depending on the amplitude of the Lorentz 
angle. The charge deflection is parallel to the vertical axis of Fig.~\ref{fig:ch_vs_pos}.
In this configuration, the detection inefficiencies related to the metal line
are unaffected by the magnetic field, while the inefficiencies due to the 
punch-through structure are suppressed. Since the overall detection inefficiency after
irradiation is dominated by the hits lost along the metal line 
the values measured with and without magnetic field generally agree.

\subsection{Charge collection across the sensor bulk~\label{sec:fluence_dep}}

As discussed in the previous sections, charge collection after heavy irradiation 
is affected by trapping of charge carriers. Electron and hole pairs produced
by traversing particles drift to the sensor electrodes where the signal is induced.
Particle irradiation produces defects in the silicon lattice which can trap these 
carriers and reduce the amplitude of the signal. 
Charges produced close to the sensor backplane and drifting for a longer distance 
are more likely to be trapped. 

\begin{figure}[hbt]
  \begin{center}
    \epsfig{file=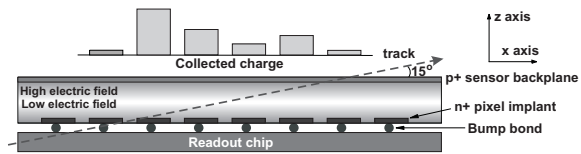,width=\linewidth}
    \caption{The grazing angle technique for determining charge collection profiles.
      The charge measured by each pixel along the $x$ direction samples 
      a different depth $z$ in the sensor.}
    \label{fig:depletion_depth} 
  \end{center}
\end{figure}
In pixel sensors, effects due to irradiation can be investigated with 
the so-called {\it grazing angle} technique~\cite{Henrich:CMSNote}. As shown in 
Figure~\ref{fig:depletion_depth} the surface of the sensor is oriented at a small
angle $\alpha=15^\circ$ with respect to the hadron beam. The charge measured by each pixel
along the $x$ direction samples a different depth $z$ in the sensor. The precise entry
point from the beam telescope is used to produce finely binned charge collection
profiles. The cluster length determines the depth over which charge is collected in the sensor.

\begin{figure}[hbt]
  \begin{center}
    \epsfig{file=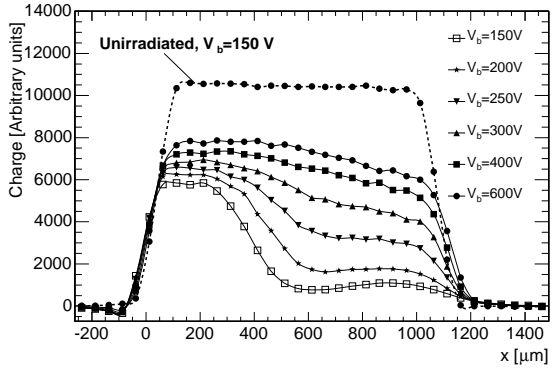,width=\linewidth}
    \caption{Charge collection profiles measured with an unirradiated (dashed line) 
      sensor and a sensor irradiated to $6\times10^{14}$~n$_{\rm eq}/{\rm cm}^{2}$ (solid line). 
      The latter is operated at bias voltages between 150~V and 600~V.~\label{fig:street_profile}}
  \end{center}
\end{figure}

The profiles measured with an unirradiated sensor and with a sensor irradiated to a fluence of 
$6\times10^{14}$~n$_{\rm eq}/{\rm cm}^{2}$ are shown in Fig.~\ref{fig:street_profile} as function of the 
distance from the beam entry point.  
The unirradiated sensor was operated at a bias voltage of 150~V which is well above
its depletion voltage (approximately 70~V). 
The profile of the collected charge is uniform and has sharp edges, indicating
full depletion. The uniformity of the signal shows
that no charge is lost when crossing the whole thickness of the sensor 
(a large $x$ coordinate corresponds a large collection distance),
as expected for an unirradiated device.  
The irradiated sensor was operated at  bias voltages between 150~V and 600~V.  
It appears to be partly depleted at 150~V, however, a second peak is observed at large
$x$. By increasing the bias voltage the amplitude of the second peak increases and
more charge is collected from the sensor side close to the backplane, while the
increase at the n+ side is about 30\%. At 600~V charge collection is saturated but
the profile is not as uniform as for the unirradiated case due to the trapping of
carriers produced far from the collecting electrode.

The observed charge collection profiles do not correspond to the classical 
picture of a partially depleted sensor and even at low bias voltages charge is collected
from the whole sensor thickness. In~\cite{Chiochia:2004qh,Chiochia:2005ag,Swartz:2005vp} it was demonstrated that the observed
profiles can be described by a two trap model producing a non-constant space charge
density across the bulk. This leads to electric field profiles with two maxima, one
at each sensor edge.
In particular, the high
electric field at the sensor backplane is necessary to explain the tails at large
$x$ observed in the charge collection profiles shown in Fig.~\ref{fig:street_profile}. 
The model is supported by measurements
of the electric field profile based on the field dependence upon electron mobility~\cite{Dorokhov:2004}.

\begin{figure}[hbt]
  \begin{center}
    \epsfig{file=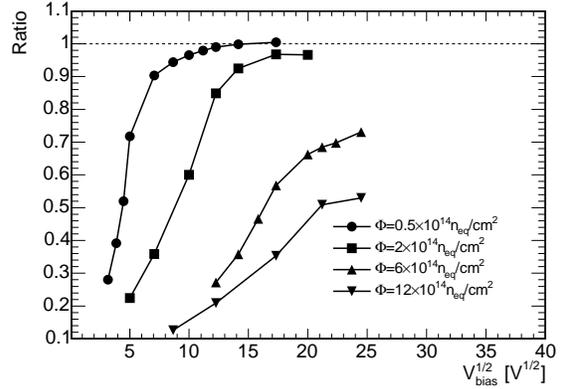,width=\linewidth}
    \caption{Ratio of the charge integrated with an unirradiated and irradiated sensors
as function of the square root of the bias voltage. The unirradiated sensor is operated at
a bias voltage of 150~V.~\label{fig:profile_ratio}}
  \end{center}
\end{figure}

In the classical picture of a type inverted device with constant space charge density
across the bulk the ratio is expected to grow as the square root of the bias voltage.
The charge collection profiles were integrated along the $x$ coordinate and the
integral measured with the irradiated sensors was divided by the value obtained
with the unirradiated sensor operated at 150~V. 
Figure~\ref{fig:profile_ratio} shows this ratio as function of the square root of the bias 
voltage. The values are not corrected for possible gain and thickness differences among the samples.
The charge ratio grows with increasing bias voltage and then reaches a plateau. 
With increasing irradiation the bias voltage at which the ratio saturates also increases,
showing that the detector has to be operated at higher bias voltage to collect 
the same deposited charge.  
The measured ratio grows faster than the $\sqrt{V}$ prediction as a 
consequence of the doubly peaked electric field profile.

\subsection{Lorentz angle~\label{sec:lorentz_angle}}
\begin{figure}[hbt]
  \begin{center}
    \epsfig{file=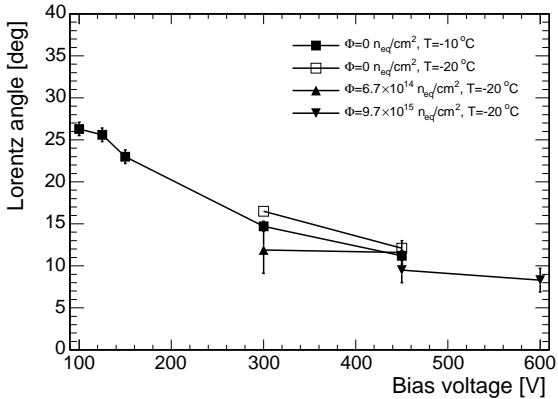,width=\linewidth}
    \caption{Lorentz angle for a 4 T magnetic field as function of bias voltage.}
    \label{fig:LA_measurement} 
  \end{center}
\end{figure}

In the presence of magnetic field the Lorentz force acts on the charge carrier, 
that are deflected from their drift along the electric field lines.
The deflection angle (Lorentz angle $\Theta_L$) can be measured with 
the grazing angle technique~\cite{Henrich:2002ed}.
\begin{figure}[hbt]
  \begin{center}
    \epsfig{file=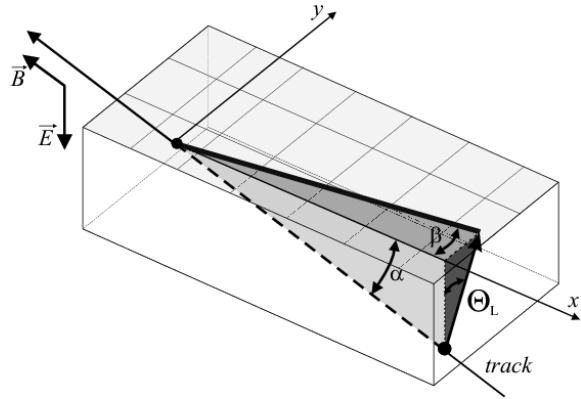,width=\linewidth}
    \caption{Measurement of the Lorentz angle with the grazing angle technique.
      The magnetic field is parallel to the beam.
    }
    \label{FIG:LA_grazing} 
  \end{center}
\end{figure}
The beam crosses the sensor at a shallow angle $\alpha=15^\circ$ and the charge carriers
are deflected by the 3 T magnetic field parallel to the beam (see Fig.~\ref{FIG:LA_grazing}). 
By measuring the angle $\beta$ between the beam direction and the direction of the
collected charge we derive the Lorentz angle
\begin{equation}
{\tan{\Theta_L} = }{\frac{\tan{\beta}}{\tan{\alpha}}}.
\end{equation}
The average cluster profile is reconstructed using the track entry point predicted 
by the beam telescope.
The cluster is sliced along the $x$ axis and the center of the each slice
is measured along the $y$ axis. The angle $\beta$ is obtained by fitting the $y$ position of 
each slice as function of $x$ with a straight line~\cite{Dorokhov:2003if}.  
A measurement without magnetic field 
is used to correct for detector misalignment with respect to the beam. 
Fig.~\ref{fig:LA_measurement} shows the Lorentz angle as function of 
bias voltage, extrapolated to 4 T magnetic field. The Lorentz angle decreases for increasing values
of the bias voltage due to the dependence of the charge carrier mobility on the electric field.
We expect the values measured at $-20^\circ$~C to be about $2^\circ$ lower due to the increase of
the electron mobility at lower temperature.

\begin{figure}[hbt]
  \begin{center}
    \epsfig{file=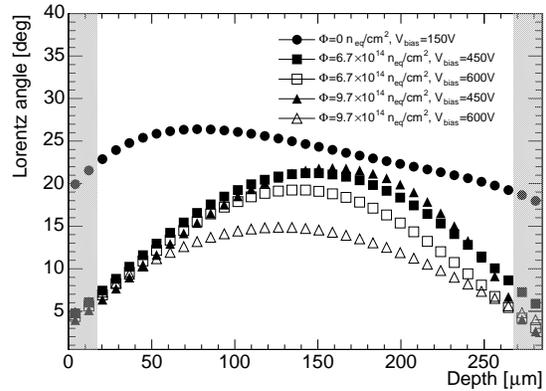,width=\linewidth}
    \caption{Lorentz angle for a 4 T magnetic field as function of sensor depth.}
    \label{fig:LA_vs_depth} 
  \end{center}
\end{figure}

A straight line fit to the charge cloud implies a constant Lorentz
deflection throughout the whole sensor thickness. A deeper analysis of 
our data sample shows that this assumption is valid only for unirradiated devices~\cite{Dorokhov:2004}.
The Lorentz angle depends on the charge carrier mobility, which is a function
of the electric field across the sensor and is given by
\begin{equation}\label{EQ:theta_l_vs_E}
 \tan \Theta_L = r_H B_x \mu(E),
\end{equation}
where $r_H$ is the Hall factor, $B_x$ is the projection of the magnetic field
along the $x$ axis and $\mu(E)$ is the carrier mobility.
In an unirradiated sensor the electric field has 
a maximum at the backplane, where the reverse bias is applied, and decreases
linearly with increasing depth. However, after irradiation,
the doubly peaked electric field produces a Lorentz angle distribution with
minima at the sensor edges and a maximum at medium depths.
The Lorentz angle as function of depth in the sensor bulk is shown in Fig.~\ref{fig:LA_vs_depth} 
for irradiated and unirradiated sensors, where a zero depth corresponds to the n+
side of the device.
The Lorentz angle is obtained from Eq.~\ref{EQ:theta_l_vs_E} and the measurement
of the carrier mobility as a function of depth~\cite{Dorokhov:2004}.

\subsection{Spatial resolution\label{sec:spatial_resolution}}

The reconstruction of the primary interaction and secondary vertices
from heavy particle decays requires a good spatial resolution.
The resolution of the pixel sensors is mainly determined by the readout pitch
and charge sharing between neighbouring cells. 
Pixels have a weak capacitive coupling
and charge sharing is mainly due to diffusion 
and drift of charge carriers under the combined effect of the magnetic and 
electric fields. After irradiation, free carriers trapping produces an 
inhomogeneous charge collection across the bulk and charge sharing between
neighbouring pixels becomes strongly nonlinear on the impact position.
In addition, the beneficial effect of the Lorentz deflection is 
reduced when a higher bias voltage is applied to provide a sufficient
drift field. In what follows we discuss measurements of the sensor
spatial resolution along the $r\phi$ direction, where the 
charge drift is affected by Lorentz deflection.

For the measurement of the spatial resolution data were recorded
with a 3 T magnetic field now perpendicular to the incoming beam. 
This corresponds to tracks in the experiment with rapidity $\eta = 0$.
Events were first selected 
with the procedure described in Section~\ref{sec:event_selection}. 
Pixels with charge above 2000 electrons were selected and clusters were formed
by adjacent pixels above threshold. Both side and corner adjacent pixels were
included in the cluster and a cluster threshold of 4500 electron was applied. 
When more than one cluster was found the cluster with
the highest charge was used for the hit position measurement.
To ensure that the cluster was entirely contained in the pixel sensor and
well reconstructed, clusters adjacent to bad pixels or to the sensor border
were rejected.
For the irradiated detectors the background from noisy pixels was reduced
by searching for further clusters in a fiducial region defined by a $5\times5$ pixels
window centered at the hit position predicted by the beam telescope.

\begin{figure}[hbt]
  \begin{center}
    \epsfig{file=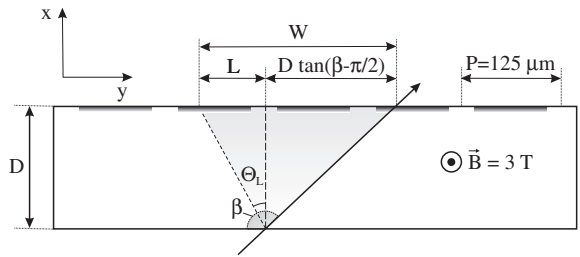,width=\linewidth}
    \caption{Determination of the impact position in the transverse plane.}
    \label{fig:pos_resolution} 
  \end{center}
\end{figure}
\begin{figure}[hbt]
  \begin{center}
    \mbox{
      \subfigure[]{\scalebox{0.90}{
	  \epsfig{file=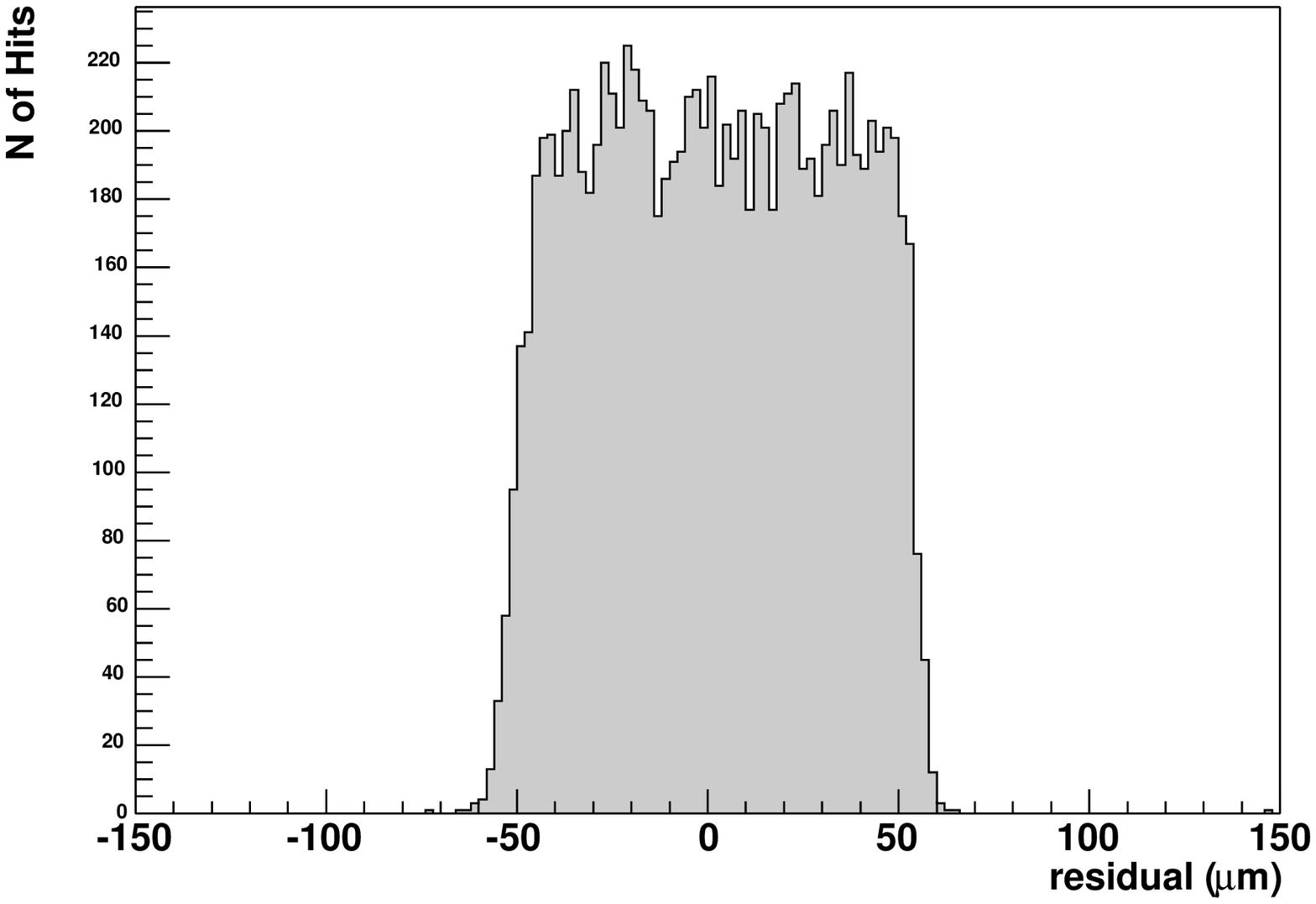,width=\linewidth}
	  \label{fig:residual_1pix} 
      }}
    }
    \mbox{
      \subfigure[]{\scalebox{0.90}{
	  \epsfig{file= 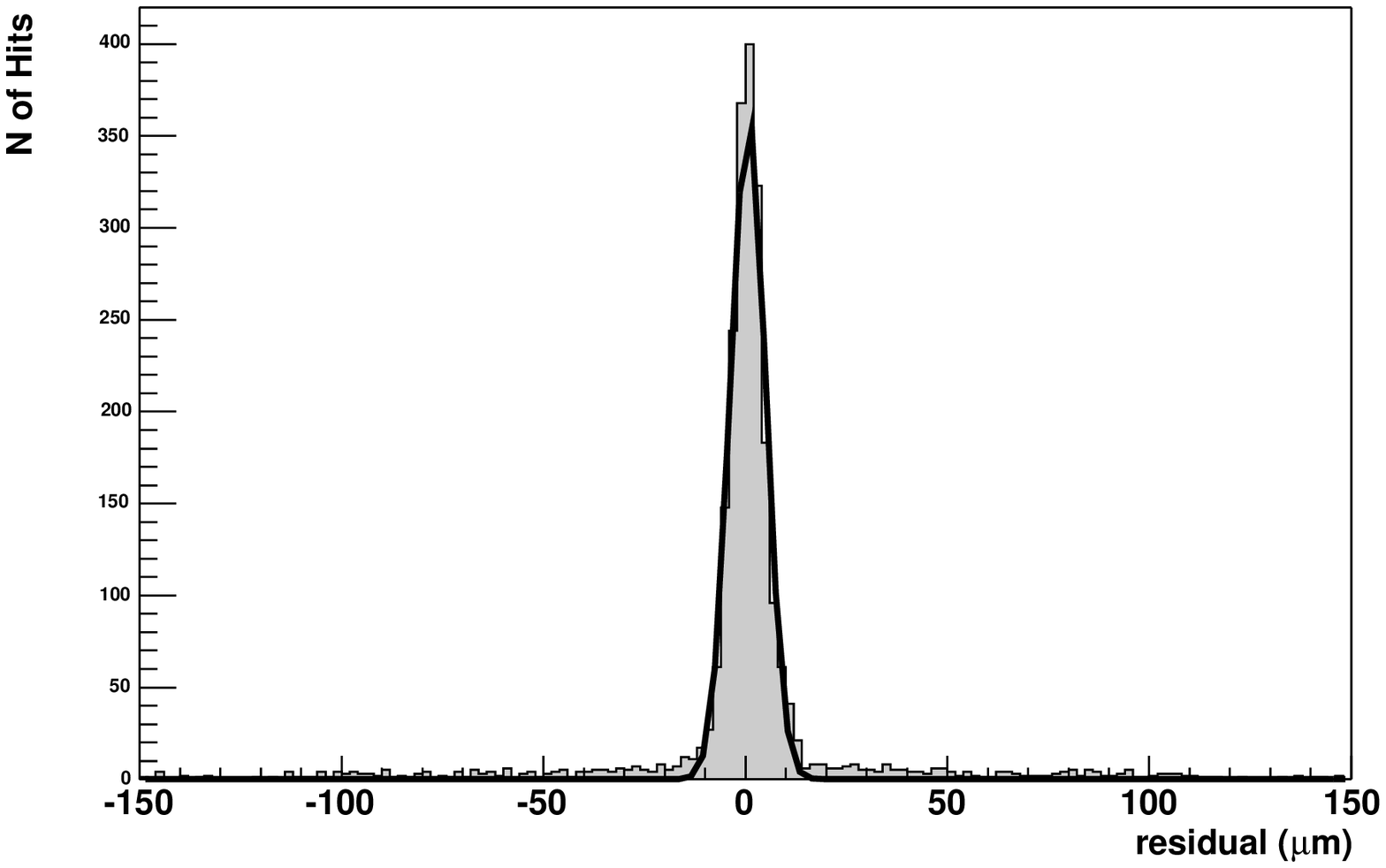,width=\linewidth}
	  \label{fig:residual_2pix} 
      }}
    }
    \caption{Residual distributions measured with an unirradiated sensor, for an impact angle
    $\beta=70^\circ$ and $V_b=150$~V. The data sample is divided in clusters of unit size (a)
      and for larger clusters (b). The solid line is a Gaussian fit.}
  \end{center}
\end{figure}

Figure~\ref{fig:pos_resolution} shows the definition of the track impact angle $\beta$
with respect to the sensor plane along the $y$ direction. The track is orthogonal
to the sensor plane (parallel to the $x$ axis). 
The magnetic field in the $z$ direction produces a 
Lorentz shift $L=D \tan \Theta_L$ toward the left direction (see Fig.\ref{fig:pos_resolution}), 
thus, the total charge width is given by
\begin{equation}
W = D \tan(\beta-\frac{\pi}{2}) + L.
\end{equation}
The cluster is projected along the $y$ direction by summing the charge collected
in the pixels with the same $y$ coordinate. For a cluster size of one
the position is given by the center of the hit pixel. For larger clusters 
the hit position is determined by the expression
\begin{equation}
y  = y_C + \frac{Q_{last}-Q_{first}}{2(Q_{last}+Q_{first})}|W|,
\label{eq:pos_reconstruction}\end{equation}
where $Q_{first}$ ($Q_{last}$) is the charge collected in the first (last) pixel
above threshold and $y_C$ is the geo\-me\-tri\-cal center of the cluster. 
The position resolution is obtained by comparing the reconstructed position
with the prediction from the beam telescope, as explained in Section~\ref{sec:alignment}.
 We define as {\it residual} the difference between reconstructed and predicted position.
For single pixel clusters the distribution is binary, as shown in Fig.~\ref{fig:residual_1pix}.
The residual distribution for clusters larger than one is shown in Fig.~\ref{fig:residual_2pix}. 
It is approximately Gaussian with non-Gaussian tails which can
be due to events where secondary electrons (or $\delta$-electrons) are produced in addition 
to the electron-hole pairs.
Scattered electrons can travel in the silicon lattice producing larger clusters, 
for which the reconstructed position is systematically displaced. 
Hereafter, the spatial resolution is given by the r.m.s.
of the residual distribution for events with cluster size equal to one and by the
$\sigma$ of a Gaussian fit for larger clusters.  

\begin{figure}[hbt]
  \begin{center}
    \mbox{
      \subfigure[]{\scalebox{0.90}{
	  \epsfig{file=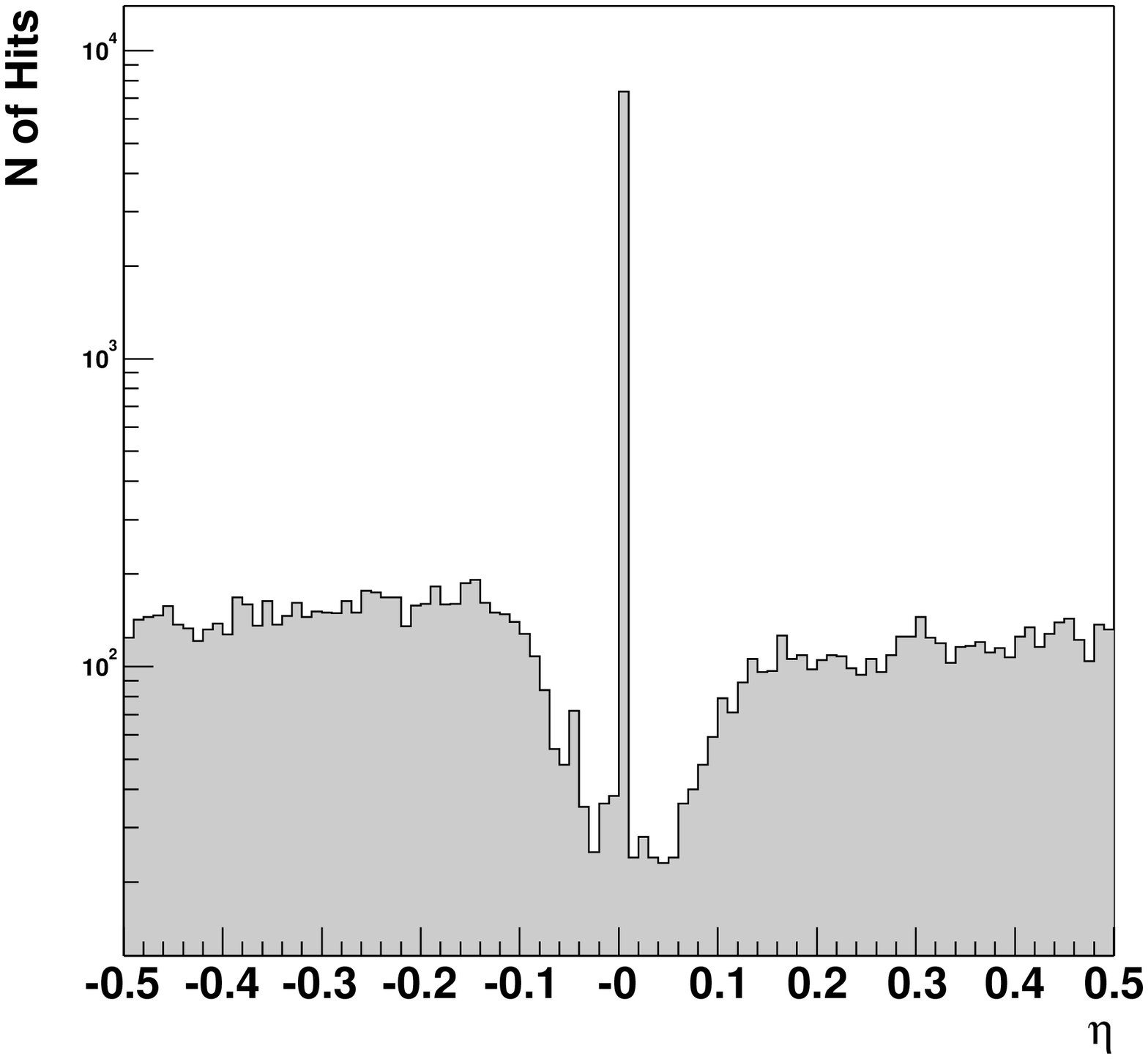,width=\linewidth}
	  \label{fig:eta_distribution} 
      }}
    }
    \mbox{
      \subfigure[]{\scalebox{0.90}{
	  \epsfig{file=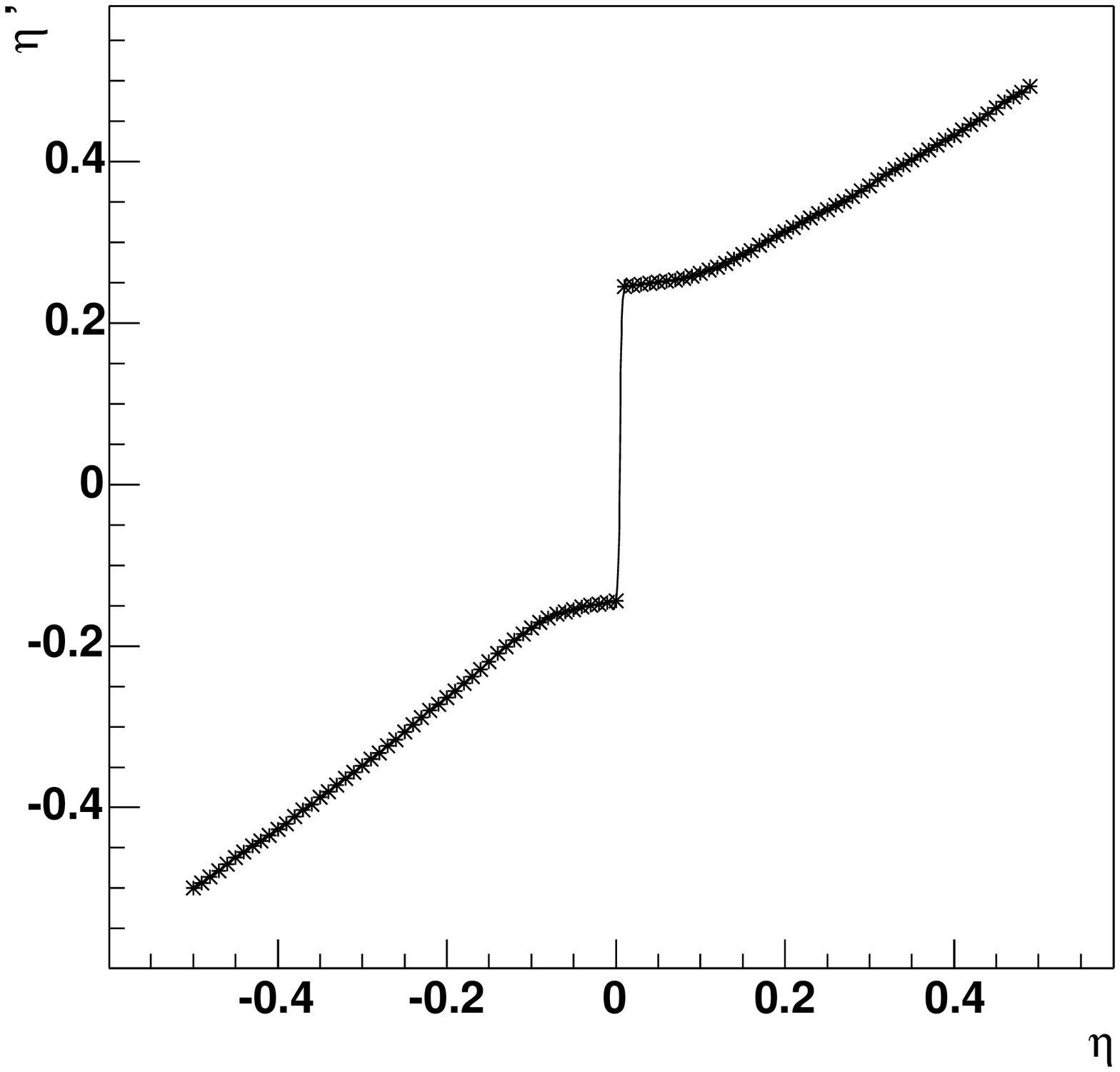,width=\linewidth}
	  \label{fig:eta_function} 
      }}
    }
    \caption{(a) Distribution of $\eta$  for a sensor irradiated to a fluence
    of $6\times10^{14}$~n$_{\rm eq}/{\rm cm}^{2}$ with $V_b=400$~V and $\beta=100^\circ$. 
    (b) Correction function $\eta'$ as function of $\eta$.}
  \end{center}
\end{figure}

To further improve the spatial resolution for events in which charge is shared
among several pixels the so-called $\eta$-correction is used~\cite{Belau:1983eh}.
As we will show, the correction is particularly effective on irradiated devices, where 
the effects of inhomogeneous charge collection are larger.
Assuming that the number of particles crossing the sensor is uniformely distributed along $y$
one expects that the reconstructed position within a pixel is also uniformely distributed.
We define $\eta$ as the non-integer
part of the position reconstructed from Eq.~\ref{eq:pos_reconstruction}. 
Figure~\ref{fig:eta_distribution} shows the $\eta$-distribution 
for all events, where $\eta=0$ corresponds to the center of the pixel cell
and $\eta= \pm 0.5$ to its borders. 
The measured distribution is almost flat in the pixel regions closer to the
pixel borders and it shows a dip at the center. The peak around zero is due
to clusters of unit size. 
The $\eta$-correction consists in rescaling the reconstructed
position in order to obtain a flat hit distribution. For each $\eta$ we associate
a corrected value given by the function
\begin{equation}
  \eta'(\eta) = \frac{\int^\eta_{-0.5}(dN/d\bar{\eta})~d\bar{\eta}}{\int^{0.5}_{-0.5}(dN/d\bar{\eta})~d\bar{\eta}} -0.5
\end{equation}
where $\eta$ is in pixel units. The $\eta'$ function is shown in Fig.~\ref{fig:eta_function}.

\begin{figure*}[hbt]
  \begin{center}
    \epsfig{file=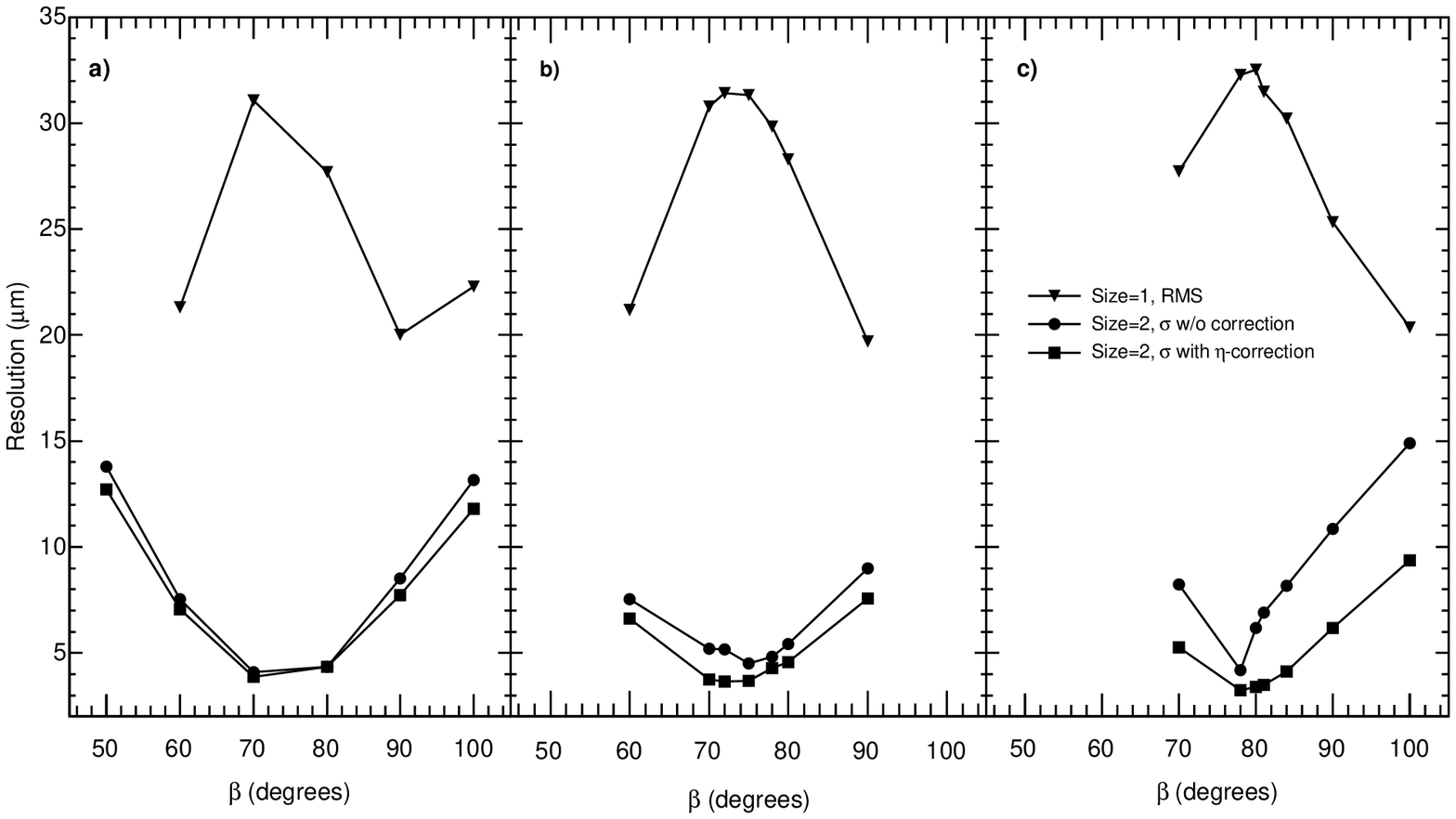,width=\linewidth}
    \epsfig{file=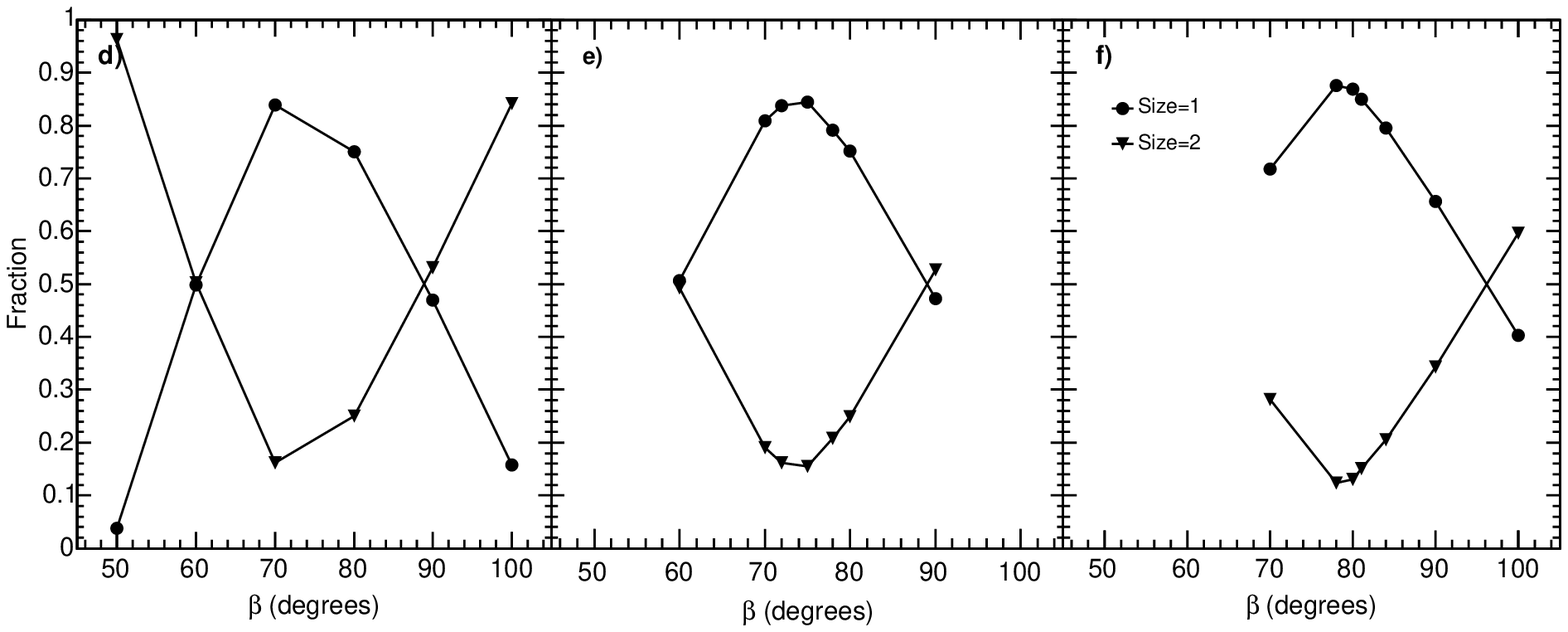,width=\linewidth}
    \caption{Top: Position resolution as function of the impact angle $\beta$ for different
      irradiation fluences: a)$\Phi = 0$ $\rm{n}_{\rm{eq}}/\rm{cm}^2$, 
      b) $\Phi = 2\times10^{14}$ $\rm{n}_{\rm{eq}}/\rm{cm}^2$, 
      c) $\Phi = 6\times10^{14}$ $\rm{n}_{\rm{eq}}/\rm{cm}^2$.
      The impact position is calculated without (solid dots) and
      with $\eta$-correction (solid squares).
      Bottom: Fraction of clusters of size one (dots) and
      size two (triangles) as function of the impact angle $\beta$ for different
      irradiation fluences: d) $\Phi = 0$ $\rm{n}_{\rm{eq}}/\rm{cm}^2$, 
      e) $\Phi = 2\times10^{14}$ $\rm{n}_{\rm{eq}}/\rm{cm}^2$, 
      f) $\Phi = 6\times10^{14}$ $\rm{n}_{\rm{eq}}/\rm{cm}^2$.
}
    \label{fig:res_vs_angle} 
  \end{center}
\end{figure*}

The position resolution is shown in Fig.~\ref{fig:res_vs_angle}(a-c)
as function of the impact angle $\beta$ and for different irradiation fluences. 
The best resolution for clusters of size bigger than one is achieved 
at $\beta\simeq \pi/2 - \Theta_L$,
for which the track direction is parallel to the Lorentz drift direction. In this case the charge width $W$
in Eq.~\ref{eq:pos_reconstruction} has a minimum and charge sharing can occur only when
the particle crosses the region between neighbouring pixels. The width of this region
is only approximately 20 $\mu$m and the position can be reconstructed with about 4 $\mu$m
precision. However, these events constitute only about 10\% of the event sample,
which is dominated by single pixel clusters. The position of the resolution minimum 
is not constant and depends on the bias voltage applied to the sensor. 
The position resolution measured with 
the $\eta$-correction is also shown in Fig.~\ref{fig:res_vs_angle}(a-c).
With increasing irradiation, trapping of charge carriers produces nonlinear charge
sharing between pixels which can be largely corrected with this method. 
In particular, larger corrections are observed for the wide charge distributions
obtained when the track is not parallel to the Lorentz shift direction. After
a fluence of $6\times10^{14}$~n$_{\rm eq}/{\rm cm}^{2}$ and for straight tracks
the precision on the impact position can be improved by a
factor two when the $\eta$-correction is applied.

Figure~\ref{fig:res_vs_angle}(a-c) also shows the position resolution for single pixel clusters.
In this case, the RMS of the residual distribution is equal to the width of
the pixel cell region in which there is no charge sharing, divided by $\sqrt{12}$. 
Since the  region has a maximum width for $\beta\simeq \pi/2 - \Theta_L$
at this value the residual width is also maximal. As for clusters of
larger size, the position of the maximum depends on the bias voltage.
A larger bias voltage produces a smaller Lorentz angle, therefore the
maximum is shifted towards larger $\beta$. For higher or lower 
values of $\beta$ the residual width is smaller and the fraction of
hits with charge sharing increases. 

The fraction of clusters of size one and two is shown in Fig.~\ref{fig:res_vs_angle}(d-f)
as function of impact angle. For $\beta=90^\circ$ the fraction
of clusters with more than one pixel is around 50\% before irradiation,
decreasing to 35\% at $6\times10^{14}$~n$_{\rm eq}/{\rm cm}^{2}$.

In the final CMS pixel detector sensors with a readout pitch of $100\times150~\mu$m$^2$ will
be used. In addition, the CMS magnetic field will be of 4 T.
With a smaller readout pitch along the $r\phi$ direction the fraction of hits
with charge sharing is expected to increase with a consequent
improvement of the overall position resolution. In absence of beam test measurements,
the expected position resolution can be estimated with a sensor simulation
implementing a model of radiation damage~\cite{physics0512027,physics0603192}. 
The model includes two defect levels with opposite charge states and trapping
of charge carriers. The simulation shows that a position resolution below 
15$~\mu$m along the CMS $r\phi$ plane can be achieved after an irradiation of
$6\times10^{14}$~n$_{\rm eq}/{\rm cm}^{2}$.
%
\section{Conclusions\label{sec:conclusions}}

The performance of the silicon sensors for the CMS barrel pixel detector were 
investigated in beam test measurements performed at the H2 beam line of the
CERN SPS. The pixel cell size of the test sensors was 125$\times$125 $\mu$m$^2$.
We expect the basic properties of the final sensor design
(100$\times$150 $\mu$m$^2$ cell size) to be similar except for the spatial
resolution.

The main results presented in this paper can be summarized as follows:
\begin{itemize}
\item A signal over noise ratio S/N$\simeq$ 65 was achieved with unirradiated sensors.
      At the specified fluence $\Phi = 6\times10^{14}$ $\rm{n}_{\rm{eq}}/\rm{cm}^2$ the
      S/N ratio is ~50.
\item For unirradiated sensors charge is uniformly collected within the pixel
      implant with exception of the punch-through structure where the collected
      signal is ~50\% lower. However, the effect of the punch-through structure
      on the overall sensor performance is very limited as it represents only
      the 2-3\% of the total surface. In the CMS experiment the effect will
      be further reduced by charge deflection due to the magnetic field.
\item The hit detection inefficiency is $\simeq 0.1\%$  for unirradiated sensors
      and applying a threshold of 2000 electrons. After a fluence of 
      $\Phi = 6\times10^{14}$ $\rm{n}_{\rm{eq}}/\rm{cm}^2$ the inefficiency
      is below 2\% and is still acceptable for the operation within CMS.
\item The ratio of the charge collected by irradiated sensors divided by the
      charge collected by an unirradiated sensor operated at 150~V was measured
      as function of the bias voltage using the grazing angle technique. 
      For a sensor irradiated at $\Phi = 6\times10^{14}$ $\rm{n}_{\rm{eq}}/\rm{cm}^2$ 
      this ratio is larger than 60\% for V$_{\rm bias}>400$ V.
\item The Lorentz angle for an unirradiated sensor and a 4 T magnetic field
      is $\simeq 24^\circ$ at 150~V and decreases at higher bias voltages. 
      After irradiation we observe 
      a Lorentz angle distribution across the sensor thickness with minima at 
      the sensor edges and a maximum at medium depths which is consistent with
      a doubly peaked electric field.
\item The position resolution for perpendicular tracks and clusters of two pixels
      is $\sigma \simeq 7~\mu$m along the coordinate with Lorentz shift. This precision
      can be achieved also after irradiation by applying the $\eta$-correction to
      reconstructed position. For clusters of a single pixel the residual distribution
      has RMS$\simeq 20~\mu$m for unirradiated sensors and  
      $\simeq 25~\mu$m for $\Phi = 6\times10^{14}$ $\rm{n}_{\rm{eq}}/\rm{cm}^2$.
\end{itemize}


\section*{Acknowledgments}

We gratefully acknowledge the contribution from Silvan Streuli (ETH Z\"urich) and Fredy Glaus (PSI) for
their immense effort with bump bonding, Federico Ravotti, Maurice Glaser and Michael Moll (CERN) for
carrying out the irradiation, Kurt B\"osiger (Z\"urich University) for the mechanical
construction, Gy\"orgy Bencze and Pascal Petiot (CERN) for the H2 beam line support
and, finally, the whole CERN-SPS team.



\bibliographystyle{elsart-num}    


\bibliography{refs}             

\begin{thebibliography}{10}
\expandafter\ifx\csname url\endcsname\relax
  \def\url#1{\texttt{#1}}\fi
\expandafter\ifx\csname urlprefix\endcsname\relax\def\urlprefix{URL }\fi

\bibitem{CMSTrackerTDR:1998}
{The~CMS~Collaboration}, {CMS} {T}racker, Technical Design Report {LHCC} 98-6,
  {CERN}, Geneva, Switzerland (1998).

\bibitem{Lindstrom:2001ww}
G.~Lindstrom, et~al., Radiation hard silicon detectors developments by the
  {RD48} ({ROSE}) {C}ollaboration, Nucl. Instrum. Meth. A466 (2001) 308--326.

\bibitem{Li:1992}
Z.~Li, H.~Kraner, Fast neutron radiation effects in silicon detectors
  fabricated by thermal oxidation processes, IEEE Trans. Nucl. Sci 39 (1992)
  577--583.

\bibitem{Beattie:1998fw}
L.~J. Beattie, et~al., The electric field in irradiated silicon detectors,
  Nucl. Instrum. Meth. A418 (1998) 314--321.

\bibitem{Casse:1998hy}
G.~Casse, E.~Grigoriev, F.~Lemeilleur, M.~Glaser, Study of evolution of active
  volume in irradiated silicon detectors, Nucl. Instrum. Meth. A426 (1999)
  140--146.

\bibitem{Eremin:2002wq}
V.~Eremin, E.~Verbitskaya, Z.~Li, The origin of double peak electric field
  distribution in heavily irradiated silicon detectors, Nucl. Instrum. Meth.
  A476 (2002) 556--564.

\bibitem{Verbitskaya:2003eg}
E.~Verbitskaya, et~al., The effect of charge collection recovery in silicon p-n
  junction detectors irradiated by different particles, Nucl. Instrum. Meth.
  A514 (2003) 47--61.

\bibitem{Chiochia:2004qh}
V.~Chiochia, et~al., Simulation of heavily irradiated silicon pixel sensors and
  comparison with test beam measurements, IEEE Trans. Nucl. Sci. 52 (2005)
  1067--1075.

\bibitem{Chiochia:2005ag}
V.~Chiochia, et~al., A double junction model of irradiated silicon pixel
  sensors for {LHC}, Nucl. Instrum. Meth. A568 (2006) 51--55.

\bibitem{Swartz:2005vp}
M.~Swartz, et~al., Observation, modeling, and temperature dependence of doubly
  peaked electric fields in irradiated silicon pixel sensors, Nucl. Instrum.
  Meth. A565 (2006) 212--220.

\bibitem{Bolla:2001ra}
G.~Bolla, et~al., Design and test of pixel sensors for the cms experiment,
  Nucl. Instrum. Meth. A461 (2001) 182--184.

\bibitem{Kaufmann:2001}
R.~Kaufmann, Development of radiation hard pixel sensors for the {CMS}
  experiment, Ph.D. thesis, Universit\"at Zurich (2001).

\bibitem{Bolla:2002my}
G.~Bolla, et~al., Sensor development for the cms pixel detector, Nucl. Instrum.
  Meth. A485 (2002) 89--99.

\bibitem{Bolla:2003si}
G.~Bolla, et~al., Irradiation studies of silicon pixel detectors for cms, Nucl.
  Instrum. Meth. A501 (2003) 160--163.

\bibitem{Dorokhov:2003if}
A.~Dorokhov, et~al., Tests of silicon sensors for the cms pixel detector, Nucl.
  Instrum. Meth. A530 (2004) 71--76.

\bibitem{Rohe:2004cm}
T.~Rohe, et~al., Position dependence of charge collection in prototype sensors
  for the cms pixel detector, IEEE Trans. Nucl. Sci. 51 (2004) 1150--1157.

\bibitem{bolla01}
G.~Bolla, et~al., Design and test of pixel sensors for the cms experiment,
  Nucl. Instrum. Methods A 461 (2001) 182--184.

\bibitem{mod-pat}
J.~Kemmer, et~al., Streifendetektor, Patent DE 19620081 A1.

\bibitem{ieee03}
T.~Rohe, et~al., Position dependence of charge collection in prototype sensors
  for the {CMS} pixel detector, IEEE Trans Nucl Sci 51 (3) (2004) 1150--1157.

\bibitem{rar96}
R.~H. Richter, L.~Andricek, T.~Gebhart, D.~Hauff, J.~Kemmer, G.~Lutz, et~al.,
  Strip detector design for {ATLAS} and {HERA-B} using two-dimensional device
  simulation, Nucl. Instrum. Methods A 377 (1996) 412--421.

\bibitem{rose}
G.~{Lindstr\"om}, M.~Ahmed, S.~Albergo, P.~Allport, D.~Anderson, L.~Andricek,
  et~al., Radiation hard silicon detectors -- developments by the {RD48}
  ({ROSE}) collaboration, Nucl. Instrum. Methods A 466 (2001) 308--326.

\bibitem{bischoff}
A.~Bischoff, N.~Findeis, D.~Hauff, P.~Holl, J.~Kemmer, P.~Klein, et~al.,
  Breakdown protection and long-term stabilisation for {S}i-detectors, Nucl.
  Instrum. Methods A 326 (1993) 27--37.

\bibitem{avset}
B.~S. Avset, L.~Evensen, The effect of metal field plates on multiguard
  structures with floating p$^+$ guard rings, Nucl. Instrum. Methods A 377
  (1996) 397--403.

\bibitem{moll}
M.~Moll, E.~Fretwurst, G.~{Lindstr\"om}, Leakage current of hadron irradiated
  silicon detectors – material dependence, Nucl. Instrum. Methods A 426
  (1999) 87--93.

\bibitem{Amsler:2002ta}
C.~Amsler, et~al., A high resolution silicon beam telescope, Nucl. Instrum.
  Meth. A480 (2002) 501--507.

\bibitem{Meer:2000}
D.~Meer, Bau und messen eines multichip pixelmodules als prototyp f\"ur den
  {CMS}-tracker, Master's thesis, Eidgen\"ossische Technische Hochschule,
  Zurich (2000).

\bibitem{Dorokhov:2005}
A.~Dorokhov, Performance of radiation hard pixel sensors for the {CMS} pixel
  detector, Ph.D. thesis, University of Z\"urich (2005).

\bibitem{Belau:1983eh}
E.~Belau, et~al., The charge collection in silicon strip detectors, Nucl.
  Instr. Meth. 214 (1983) 253.

\bibitem{Lari:2002wb}
T.~Lari, Test beam results of {ATLAS} pixel sensors, hep-ex/0210045 (2002).

\bibitem{Rohe:2004}
T.~Rohe, et~al., Fluence dependence of charge collection of irradiated pixel
  sensors, Nucl. Instrum. Meth. A552 (2005) 232--238.

\bibitem{Henrich:CMSNote}
B.~Henrich, et~al., Depth profile of signal charge collected in heavily
  irradiated silicon pixels, CMS Note 1997/021 (1997).

\bibitem{Dorokhov:2004}
A.~Dorokhov, et~al., Electric field measurement in heavily irradiated pixel
  sensors, Nucl. Instrum. Meth. A560 (2006) 112--117.

\bibitem{Henrich:2002ed}
B.~Henrich, R.~Kaufmann, Lorentz-angle in irradiated silicon, Nucl. Instrum.
  Meth. A477 (2002) 304--307.

\bibitem{physics0512027}
E.~Alagoz, V.~Chiochia, M.~Swartz, Simulation and hit reconstruction of
  irradiated pixel sensors for the {CMS} experiment, Nucl. Instrum. Meth. A566
  (2006) 40--44.

\bibitem{physics0603192}
V.~Chiochia, E.~Alagoz, M.~Swartz, Sensor simulation and position calibration
  for the {CMS} pixel detector, Nucl. Instrum. Meth. A569 (2006) 132--135.

\end{thebibliography}

\end{document}